\documentclass{aastex701}

\usepackage{adjustbox}
\usepackage{rotating}
\usepackage{makecell}
\usepackage{multirow}
\usepackage{booktabs}
\usepackage{amsmath}
\usepackage{float}
\usepackage{soul}

\hypersetup{linkcolor=red,citecolor=blue,filecolor=cyan,urlcolor=magenta}

\begin{document}

\title{BSN-VIII: Detailed Photometric Modeling of Ten W UMa Contact Binaries\\
and a Revised Empirical Period-Mass Relationship}

\author[0000-0002-0196-9732]{Atila Poro}
\altaffiliation{atila.poro@obspm.fr}
\affiliation{LUX, Observatoire de Paris, CNRS, PSL, 61 Avenue de l'Observatoire, 75014 Paris, France}
\affiliation{Astronomy Department of the Raderon AI Lab., BC., Burnaby, Canada}
\email{atila.poro@obspm.fr}

\author[0000-0002-9490-2093]{S. Javad Jafarzadeh}
\affiliation{Department of Physics, University of Texas at Dallas, 800 W. Campbell Rd., Richardson, TX 75080, USA}
\email{sj.jafarzade@gmail.com}

\author[0000-0003-3590-335X]{Kai Li}
\affiliation{Shandong Key Laboratory of Space Environment and Exploration Technology, Institute of Space Sciences, School of Space Science and Technology, Shandong University, Shandong 264209, People’s Republic of China}
\email{kaili@sdu.edu.cn}

\author[0000-0003-0354-8568]{Jean-François Coliac}
\affiliation{Double Stars Committee, Société Astronomique de France, 75016 Paris, France}
\email{jfcoliac@gmail.com}

\author[0009-0004-8426-4114]{Sabrina Baudart}
\affiliation{Double Stars Committee, Société Astronomique de France, 75016 Paris, France}
\email{sabrina.baudart@gmail.com}

\author{Meng Guo}
\affiliation{Shandong Key Laboratory of Space Environment and Exploration Technology, Institute of Space Sciences, School of Space Science and Technology, Shandong University, Shandong 264209, People’s Republic of China}
\email{1728609732@qq.com}

\author[0009-0008-3298-4194]{Patrick Wullaert}
\affiliation{Double Stars Committee, Société Astronomique de France, 75016 Paris, France}
\email{patrick.wullaert@saf-astronomie.fr}

\author{David Valls-Gabaud}
\affiliation{LUX, Observatoire de Paris, CNRS, PSL, 61 Avenue de l'Observatoire, 75014 Paris, France}
\email{david.valls-gabaud@obspm.fr}

\author{Anis Ben Lassoued}
\affiliation{Double Stars Committee, Société Astronomique de France, 75016 Paris, France}
\email{anisbenlassoued68@gmail.com}

\author{Daniel Verilhac}
\affiliation{Double Stars Committee, Société Astronomique de France, 75016 Paris, France}
\email{daniel.verilhac@free.fr}

\author{Laurent Corp}
\affiliation{Double Stars Committee, Société Astronomique de France, 75016 Paris, France}
\email{laucorp@wanadoo.fr}

\author{Emmanuel Foguenne}
\affiliation{Double Stars Committee, Société Astronomique de France, 75016 Paris, France}
\email{emmanuel.foguenne@gmail.com}

\author[0000-0002-8505-378X]{Hervé Lerat}
\affiliation{Double Stars Committee, Société Astronomique de France, 75016 Paris, France}
\email{herve.lerat.1966@gmail.com}

\author[0000-0002-7901-7213]{Jean-Baptiste Marquette}
\affiliation{Double Stars Committee, Société Astronomique de France, 75016 Paris, France}
\email{jb.marquette@gmail.com}

\author{Serge Vasseur}
\affiliation{Double Stars Committee, Société Astronomique de France, 75016 Paris, France}
\email{hipparcos.astro@outlook.fr}

\author[0009-0007-2048-4865]{Anica Lekic}
\affiliation{Double Stars Committee, Société Astronomique de France, 75016 Paris, France}
\affiliation{IPSA Institut Polytechnique des Sciences Avancées, 94200 Ivry-sur-Seine, France}
\email{anica.lekic@ipsa.fr}

\author[0000-0002-1972-8400]{Fahri Alicavus}
\affiliation{Çanakkale Onsekiz Mart University, Faculty of Science, Department of Physics, 17020, Çanakkale, Türkiye}
\affiliation{Çanakkale Onsekiz Mart University, Astrophysics Research Center and Ulupnar Observatory, 17020, Çanakkale, Türkiye}
\email{fahrilcvs@gmail.com}

\author[0000-0001-9809-7493]{Neslihan Alan}
\affiliation{Fatih Sultan Mehmet Vakif University, Faculty of Humanities and Social Sciences, Department of History of Science, 34664, Istanbul, Türkiye}
\email{neslihan.alan@gmail.com}


\begin{abstract}
This study continues our ongoing research on contact binary systems by presenting a detailed analysis of 10 targets. Ground-based observations from six different observatories were conducted and used together with TESS data for the analysis process. Photometric data from our observations were reduced with the recently developed AutoWISP pipeline, yielding high-quality light curves with reliable precision for analysis. An investigation of orbital period variations identifies long-term trends in six of the ten analyzed binaries, including three that also display cyclic variations. Four targets show essentially constant orbital periods. The secular trends are attributed to mass transfer. The cyclic modulations in three systems are caused by either magnetic activity cycles or the Light-Travel Time Effect (LTTE) of a third body, while that in the remaining system is solely due to the LTTE. The BSN application was used to model the photometric light curves of the 10 target binaries. Iterative fitting and MCMC refinement provided robust estimates of the system parameters, while starspot modeling was applied for systems showing O'Connell-effect asymmetries. We refine the empirical orbital period–mass relationship for short-period contact binaries by analyzing a homogeneous dataset of systems and deriving an updated primary-mass calibration based on the spectroscopic subset. Using this calibrated relation, the fundamental parameters of the studied systems were subsequently estimated.
\end{abstract}


\section{Introduction}
Stellar systems belonging to the W Ursae Majoris (W UMa) type are close binary stars sharing a common outer envelope, which places both components in physical contact. These stars usually orbit each other in less than a day and have comparable surface temperatures, allowing direct exchange of energy. Their mutual interaction produces nearly equal eclipse depths and continuous light variations that reflect the physical connection between the components.

Contact binaries form when both stellar components fill or overfill their Roche lobes, causing their outer equipotential surfaces to merge and creating a common convective envelope where mass, energy, and angular momentum circulate freely (\citealt{1971ARA&A...9..183P,1981ApJ...245..650M}). The extent to which the stellar surfaces lie above the inner critical Roche potential provides insight into the physical configuration and evolutionary condition of the system (\citealt{1968ApJ...151.1123L,2005ApJ...629.1055Y}). The contact configuration can be characterized by the fillout factor, a dimensionless parameter that specifies the position of the common equipotential surface relative to the inner and outer critical equipotential surfaces (\citealt{1971ApJ...166..605W}). High fillout values correspond to deeper contact geometry, enhanced thermal interaction within the binary system, and more efficient exchange of energy throughout the shared envelope (\citealt{1976ApJ...205..208L}). Low fillout factor values, by contrast, reflect shallow-contact configurations that are prone to thermal imbalance and variations in contact depth. Consequently, the fillout factor functions as a central structural parameter for evaluating the geometry, thermal state, and evolutionary progression of contact binaries (\citealt{2003ASPC..293...76W}).

Orbital period variations in contact binaries commonly appear as parabolic, cyclic, or apparently linear trends, each corresponding to a distinct physical mechanism that may operate simultaneously within a single system. Parabolic variations generally indicate long-term evolutionary changes associated with mass exchange or systemic mass loss (\citealt{2018PASJ...70...90K}). The curvature of the orbital period variation diagram reflects whether material is moving from the less massive to the more massive component. Linear trends can result when the available observational baseline is insufficient to detect significant orbital period variations. During long-term orbital period variations, some systems exhibit pronounced cyclic modulations in the orbital period variation diagrams, and several mechanisms can account for such periodic behavior (\citealt{2016A&A...587A..34V}). The most common explanation is the LTTE caused by a tertiary companion, whose gravitational influence causes the binary to follow an orbit around the common center of mass of the triple system, producing strictly periodic timing variations (\citealt{2016MNRAS.455.4136B}).
Representative observational evidence for this mechanism has been reported for many contact binaries, such as V1062 Her and V1067 Her (\citealt{2025MNRAS.537.3366Z}).
Another possible explanation is that cyclic variations can arise from magnetic cycles operating within one or both stellar components (\citealt{1998MNRAS.296..893L}).
 During these cycles, changes in the internal magnetic field redistribute angular momentum and alter the stellar quadrupole moment, producing Applegate-type (\citealt{1992ApJ...385..621A}) modulation. Additional sources of cyclic behavior include long-term magnetic cycles that can modify the stellar radii or energy output (\citealt{2010MNRAS.405.1930L}). Dynamical interactions in hierarchical triple or quadruple systems may introduce multi-periodic signatures. In rare cases, orbital precession or nodal regression can also contribute to cyclic variations (\citealt{2015MNRAS.448..946B}). The presence of long-term orbital period variations together with one or more superimposed cyclic variations is expected in some contact binaries (\citealt{2001MNRAS.328..914Q}). Complete interpretation of orbital period variations requires dense, long-term timing coverage over multiple decades, supported by high-quality observations obtained from both ground facilities and space missions.

Studies in the literature have investigated empirical correlations among the physical parameters of contact stars (e.g., \citealt{2009CoAst.159..129G,2021ApJS..254...10L,2022MNRAS.510.5315P,2024RAA....24a5002P,2026RAA....26c5022A}). These efforts aim to examine and identify the structural and evolutionary processes that govern the mutual dependencies among these parameters and shape the overall behavior of the systems. Such correlations provide independent and complementary constraints, particularly when high-quality spectroscopic or photometric observations are unavailable, enabling the estimation of fundamental physical quantities including masses, radii, temperatures, orbital period variations, and the degree of contact (\citealt{2025MNRAS.538.1427P}). They further clarify how thermal coupling and energy redistribution within the common convective envelope influence the configuration of W UMa-type binaries, and how parameters such as mass ratio, orbital period, and energy transfer collectively regulate their long-term evolution. In addition, these empirical relations support statistical population analyses and facilitate the identification of systems exhibiting deviations associated with mass-transfer episodes or magnetic activity (\citealt{2018ApJ...859..140C}). With the continuous expansion of modern large-scale surveys, systematic refinement and updating of these correlations is required to enhance our knowledge of the relationship between the structural properties and evolutionary processes of short-period contact binaries.

This study provides a photometric investigation of 10 W UMa-type eclipsing contact binaries using ground-based observations, aiming to improve our knowledge of their orbital and physical parameters. Our work builds upon previous efforts (\citealt{2025MNRAS.537.3160P, 2025AJ....170..214P,2025PASP..137h4201P,2025PASP..137k4203P,2025PASP..137l4202P, 2026PASP..138c4203S}), contributing new observational data and analyses for additional W UMa systems as part of the Binary Systems of South and North (BSN) project.

\vspace{0.6cm}
\section{Dataset}
\subsection{Target Systems}
A total of 10 eclipsing binary stars were analyzed, including Gaia DR3 1721085002866579840 (hereinafter G1721), Gaia DR3 2161226688650387200 (hereinafter G2161), NO Leo, TIC 422347573, V337 UMa, V369 Boo, V475 Ser, V640 Aur, V689 Vir, and V702 Aur. Detailed investigations of these contact binary systems have not previously been reported. Additionally, the BSN project database contains photometric observations for them, providing sufficient coverage for a reliable analysis. Table \ref{Tab:systemsinfo} provides an overview of the target systems from Gaia DR3 (\citealt{2023A&A...674A..33G}) and the TESS Input Catalog (TIC). Additionally, we determined and report the maximum $V$-band apparent magnitude for the systems in Table \ref{Tab:systemsinfo}. According to astronomical catalogs and databases, target binaries are identified as contact stars, including the All-Sky Automated Survey for Supernovae (ASAS-SN; \citealt{2014ApJ...788...48S,2018MNRAS.477.3145J}) and the Variable Star Index (VSX\footnote{\url{https://vsx.aavso.org/}}). The apparent magnitudes of the targets range from about 12.5 to 14.5, and the system temperatures reported in Gaia DR3 and the TIC lie between 4800 K and 6000 K (Table \ref{Tab:systemsinfo}). Additionally, the systems are characterized by short orbital periods, falling within the narrow interval of 0.25-0.38 days.

\begin{table*}
\renewcommand\arraystretch{1.2}
\caption{Coordinates, distance, and temperature of the system from Gaia DR3, temperature from TIC, along with the maximum apparent magnitude in the $V$ band obtained in this study.}
\centering
\begin{center}
\footnotesize
\begin{tabular}{c c c c c c c}
\hline
System & RA$.^\circ$(J2000) & Dec$.^\circ$(J2000) & $d$(pc) & $T_{\text{Gaia}}$(K) & $T_{\text{TIC}}$(K) & $V_{\text{max}}$(mag.)\\
\hline
Gaia DR3 1721085002866579840(G1721) & 235.263584 & 80.720098 & 379(2) & 5766(17) & 5310(174) & 13.06(10)\\
Gaia DR3 2161226688650387200(G2161) & 274.457483 & 64.250516 & 639(5) & 5777(34) & 5996(143) & 13.45(9)\\
NO Leo & 165.008620 & 4.701662 & 270(1) & 4855(38) & 5004(199) & 13.23(9)\\
TIC 422347573 & 197.100337 & 6.479036 & 471(73) & 5813(10) & 5922 & 12.59(10)\\
V337 UMa & 152.786839 & 50.589200 & 773(42) & 5437(10) & 5537(24) & 13.97(12)\\
V369 Boo & 220.912699 & 53.793524 & 675(6) & 5737(14) & 5609(163) & 13.70(14)\\
V475 Ser & 237.444251 & 23.914723 & 297(1) & 4887(20) & 4826(131) & 13.67(18)\\
V640 Aur & 90.531812 & 52.528766 & 388(2) & & 5641(189) & 13.09(15)\\
V689 Vir & 194.990592 & -6.466478 & 243(2) & 5156(32) & 5216(135) & 12.82(13)\\
V702 Aur & 78.713064 & 39.219093 & 925(18) & 5991(17) & 5806(30) & 14.55(12)\\
\hline
\end{tabular}
\end{center}
\label{Tab:systemsinfo}
\end{table*}

\vspace{0.6cm}
\subsection{Observations}
Six ground-based observatories contributed observations of the 10 target systems: Observatoire Astronomique de Sabichette (OASa) in France, Tunisia Sousse Observatory (TSO) in Tunisia, Observatoire Astronomique des Binaires (OABAC) in France, Observatoire SADR Poroto MPC code X03 (SADR) in Chile, Observatoire de Haute-Provence MPC code 511 (OHP) in France, and AstroKoT MPC code D99 (AstroKoT) in France. Observations were carried out over 29 nights during 2024 and 2025. Various types of telescopes, including both refractors and reflectors of different apertures, were employed throughout the observing campaign. Details of the observations are given in Table \ref{Tab:g-observations}.

Time-series photometric observations were obtained from the Transiting Exoplanet Survey Satellite (TESS). TESS uses four wide-field cameras to monitor regions of the sky for approximately 27.4 days per sector. TESS photometric data in the broad "TESS:T" passband (600–1000 nm) were retrieved for the target binary systems, except for V702 Aur, which lacks TESS time-series data. To extract minima in Section 3, all available sectors with time-series data were utilized. Light curve solution was performed using data from the TESS-Science Processing Operations Center (SPOC), and Quick-Look Pipeline (QLP), pipelines whenever available. The selection of the pipeline and observing sector was based primarily on data quality, low exposure length, and preference for one of the most recent available sectors to minimize the potential impact of long-term variability (Table \ref{Tab:tess}). For TESS-SPOC pipelines, the Simple Aperture Photometry (SAP) flux was used. A segment of approximately 1500-2200 points from the latest available TESS sector with 200-s cadence was selected for the light curve solution. This portion exhibiting the lowest scatter was identified using a Python code applied to the full time-series data. Baseline variations within the selected segment were examined, and no significant long-term trends were detected; consequently, normalization and detrending of the selected segment were performed using the Lightkurve package to remove short-term variations while preserving the segment's integrity. It should be noted that, to reduce the scatter in the light curve, consecutive data points were grouped into bins of three to five points, with the bin size chosen according to the local scatter of the light curve segment. We also compared the obtained light curve with other TESS sectors for each target. Moreover, since this study is not limited to single-band data, ground-based observations were also compared with the TESS light curve during modeling. The corresponding TESS sectors are given in Table \ref{Tab:tess}. The TESS observations analyzed in this work were downloaded from the Mikulski Archive for Space Telescopes (MAST).

\begin{table*}
\renewcommand\arraystretch{1.2}
\caption{Observing information for the target systems, obtained at various BSN observatories.}
\centering
\begin{center}
\footnotesize
\begin{tabular}{c c c c c c}
\hline
Target & Observatory & Telescope & CCD & Observation & Filter and\\

System & Site & Aperture(mm) & Model & Date & Exposure time(s)\\
\hline
G1721	&	OASa	&	Refractor 102	&	ASI 1600 MM Pro (ZWO)	&	29/04/2025	&	$V$(150)	\\
G2161	&	OASa	&	Refractor 102	&	ASI 1600 MM Pro (ZWO)	&	30/04/2025	&	$V$(150)	\\
NO Leo	&	SADR	&	Reflector 360	&	CMOS PlayerOne IMX571 Mono	&	13\&14/04/2025	&	$R$-Baader(120)	\\
TIC 422347573	&	TSO	&	Reflector 150	&	ASI 533 MM Pro (ZWO)	&	26\&27\&28/05/2025	&	$R_c$(30)	\\
V337 UMa	&	OABAC	&	Reflector 150	&	ASI ZWO 183mm Pro	&	06\&07/02/2025, 17\&18/03/2025	&	$V$(180), $R_c$(180)	\\
V369 Boo	&	OABAC	&	Reflector 150	&	ASI ZWO 183mm Pro	&	21\&22\&23/04/2025	&	$V$(180)	\\
V369 Boo	&	OHP	&	Reflector 800	&	SBIG STXL 6303E	&	05/04/25	&	$r'$(60)	\\
V475 Ser	&	AstroKoT	&	Reflector 250	&	ASI ZWO 183mm Pro	&	16\&17\&18/05/25	&	$B$(420), $R_c$(420)	\\
V640 Aur	&	OABAC	&	Reflector 150	&	ASI ZWO 183mm Pro	&	29\&30/12/2024, 13\&14/01/2025	&	$V$(180), $R_c$(180)	\\
V689 Vir	&	SADR	&	Reflector 360	&	CMOS PlayerOne IMX571 Mono	&	27\&28/04/2025	&	$V$(180), $R$-Baader(180)	\\
V702 Aur	&	OABAC	&	Reflector 150	&	ASI ZWO 183mm Pro	&	27\&28\&29/12/2024	&	$V$(180), $R_c$(180)	\\
\hline
\end{tabular}
\end{center}
\label{Tab:g-observations}
\end{table*}

\begin{sidewaystable*}
\centering
\footnotesize
\setlength\tabcolsep{3pt}
\caption{TESS observation specifications for target systems. These time-series data were utilized for orbital period variation analysis via minima extraction, and photometric light curve modeling in the investigation. Exposure Length (E.L.) is indicated in the table.}
\begin{tabular}{cc|ccc|c}
\hline
System & TIC & Sector for O-C & Available E.L.(s) & Observation Year & Sector for modeling/E.L.(s)/Pipeline\\
\hline
G1721 & 159343873 & 14, 15, 19, 20, 21, 25, 26, 40, 41, 47, 48, & 200, 600, 1800 & 2019, 2020, 2021, 2022, 2023, 2024 & 75/200/TESS-SPOC\\
      &           & 52, 53, 59, 60, 73, 74, 75, 79, 86 &                &                        & \\
G2161 & 233152163 & 15, 16, 17, 18, 19, 20, 22, 23, 24, 25, 26, 40, & 200, 600, 1800 & 2019, 2020, 2021, 2022, 2023, 2024 & 85/200/QLP\\
      &           & 47, 48, 49, 50, 52, 53, 54, 55, 56, 57, 58, 60, &                &                        & \\
      &           & 73, 74, 75, 76, 77, 78, 79, 80, 81, 83, 84, 85 &                &                        & \\
NO Leo        & 374367649 & 45, 46, 72 & 200, 600 & 2021, 2023 & 72/200/TESS-SPOC\\
TIC 422347573 & 422347573 & 23, 50 & 600, 1800 & 2020, 2022 & 23/1800/QLP-50/600/QLP\\
V337 UMa      & 371520599 & 48, 75 & 600 & 2022, 2024 & 48/600/TESS-SPOC\\
V369 Boo      & 161001574 & 16, 22, 23, 49, 50, 76, 77 & 200, 600, 1800 & 2019, 2020, 2022, 2024 & 77/200/TESS-SPOC\\
V475 Ser      & 459966504 & 24, 51, 78 & 200, 600, 1800 & 2020, 2022, 2024 & 78/200/QLP\\
V640 Aur      & 467114320 & 19, 59, 60, 73 & 200, 1800 & 2019, 2022, 2023 & 73/200/QLP\\
V689 Vir      & 319464682 & 46, 91 & 200, 600 & 2021, 2025 & 91/200/QLP\\
\hline
\end{tabular}
\label{Tab:tess}
\end{sidewaystable*}

\vspace{0.6cm}
\subsection{Data Reductions}
We performed data reduction using Automated Wide-Image Stellar Photometry (AutoWISP; \citealt{2025AJ....170..250R}), a pipeline designed to extract high quality light curves from photometric observation.

AutoWISP offers several notable advantages. It operates effectively on color images from Digital Single-Lens Reflex (DSLR) cameras and Charge-Coupled Devices (CCDs), produces high-quality light curves for all stars in the field in a single run, and minimizes different types of errors even without bias, dark, or flat frames when such calibrations are unavailable. The software supports simultaneous photometric extraction in any number of apertures, incorporates detailed models of the Point-Spread Function (PSF) and Pixel-Response Function (PRF), and is optimized for highly heterogeneous wide-field datasets.

The "automated" aspect refers to the fact that, once the observer defines a survey configuration, including telescope and camera properties, in addition to calibration, reduction, and trend-filtering parameters, no further human interaction is required. AutoWISP proceeds from raw frames to fully processed light curves for all detected sources. The only user-dependent step is the magnitude-fitting procedure, for which the user simply selects a reference frame. Telescope and camera specifications, along with all configuration parameters, may be saved to JSON files and re-used for future reductions with the same equipment. Additional details will be provided in a forthcoming publication describing the AutoWISP Browser User Interface.

The full reduction workflow includes image calibration, source detection, astrometric plate solving, stellar profile modeling, aperture photometry, PSF fitting, magnitude fitting, and light curve generation. AutoWISP also implements two post-processing algorithms, External Parameter Decorrelation (EPD) and the Trend Filtering Algorithm (TFA). While EPD and TFA can reduce systematics, they may also suppress genuine astrophysical variability, particularly the large-amplitude signals characteristic of contact binaries. After experimentation, we found that magnitude fitting provided the most reliable results for our targets; consequently, EPD and TFA were not applied in the final analysis. Detailed descriptions of each processing stage are documented in \cite{2025AJ....170..191P} and \cite{2025AJ....170..250R}.

A number of reduction parameters had to be adjusted from their default values for all observations. For example, the brightness threshold for source extraction needed to be set so that a sufficient number of stars were detected while avoiding false positives. In addition, a suitable set of magnitude fitting correction parameters and its order had to be chosen. These parameters are required to convert raw magnitudes to standard magnitudes by comparing each target frame with a pre-selected reference frame and applying the polynomial corrections. The goal is to make the corrected magnitudes in the target frame match the stellar magnitudes in the reference frame as closely as possible.

Furthermore, we tested a wide range of 20 photometric apertures, from 1.0 to 5.0 pixels in steps of 0.2 pixels. This allowed us to adopt the aperture that best preserved the intrinsic variability amplitude of the target while still providing stable photometric performance.

\begin{figure*}
\centering
\includegraphics[width=0.99\linewidth]{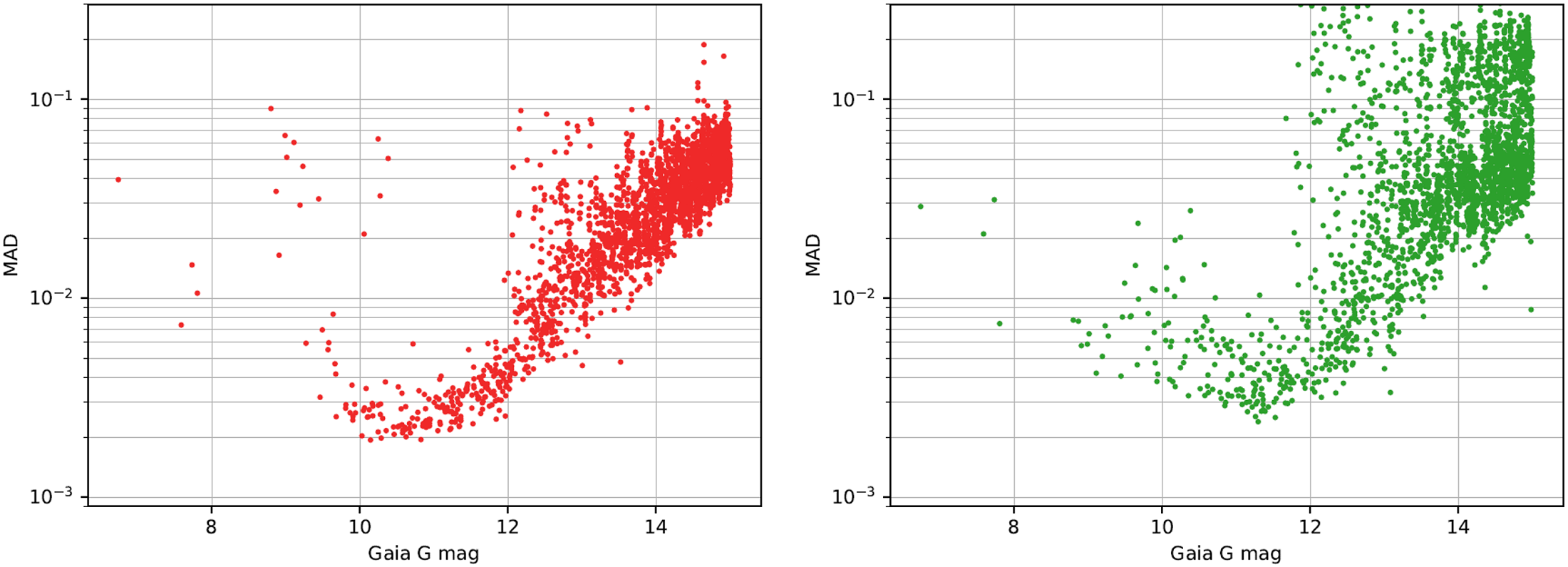}
\caption{Minimum median absolute deviation over all apertures versus Gaia $g$ magnitude for the observations of V702 Aur in two filters, with the $R$ band on the left and the $V$ band on the right. The mean magnitudes of this object in Gaia $G$, $BP$, and $RP$ bands are 14.396, 14.846, and 13.777, respectively.}
\label{Fig:v702_filters}
\end{figure*}

\begin{figure*}
\centering
\includegraphics[width=0.99\linewidth]{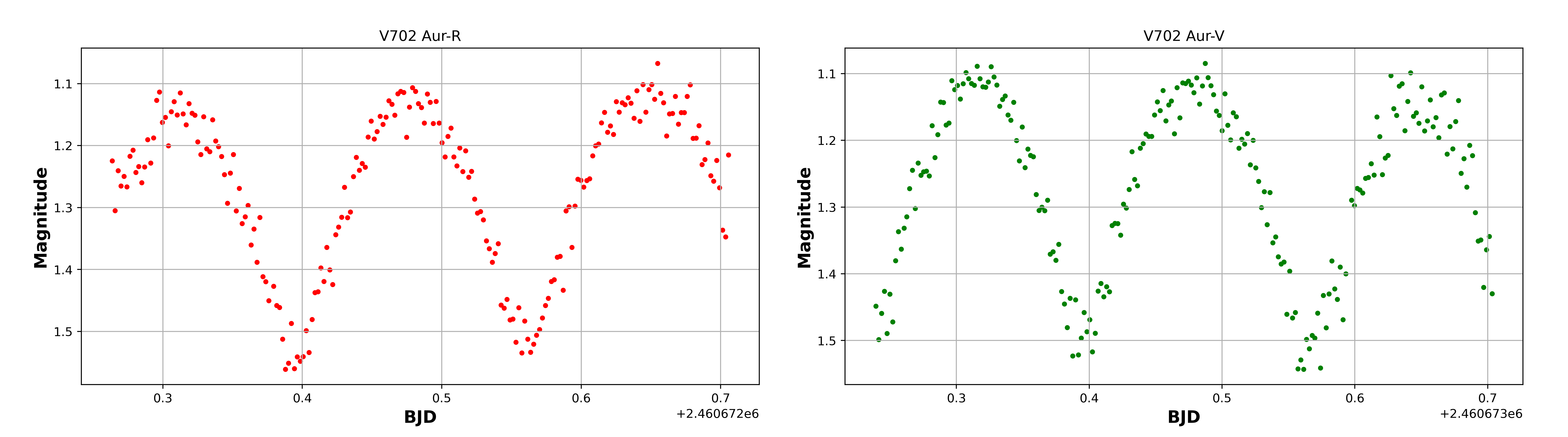}
\caption{Light curves of V702 Aur in two filters, with the $R$ band shown on the left and the $V$ band on the right. Magnitudes are plotted with an arbitrary offset.}
\label{Fig:v702_lc}
\end{figure*}

To quantify photometric precision, we computed the Median Absolute Deviation (MAD) of the measurements across all apertures for each star and plotted the minimum MAD as a function of Gaia $g$-band magnitude. Two representative examples for V702 Aur are shown in Figure \ref{Fig:v702_filters}. As expected, the MAD decreases toward brighter stars, reaching values as low as ~0.002 mag for stars with magnitudes 10-11. At even brighter magnitudes, the MAD increases due to partial saturation. The Gaia-$g$-magnitude of this target is 14.396, corresponding to a typical MAD of about 3\% for stars of similar brightness. Figure \ref{Fig:v702_lc} presents the corresponding light curves of this system.

Finally, for each object, we extracted the observation times, magnitudes, and magnitude uncertainties from the selected aperture of the AutoWISP-generated HDF5 light curve files and used these values in the subsequent scientific analysis.

\vspace{0.6cm}
\section{Investigation of Orbital Period Variations}
Possible orbital period changes in the ten target systems were evaluated by comparing the observed and calculated times of minima through the O-C method (\citealt{2013NewA...21...46L,2019RAA....19..147L,2022AJ....164..202L}). Eclipse timings were collected and extracted from photometric surveys including the ASAS-SN, the Zwicky Transient Facility (ZTF; \citealt{2019PASP..131f8003B,2019PASP..131a8003M}), the Transiting Exoplanet Survey Satellite (TESS; \citealt{2014SPIE.9143E..20R}), the Wide Angle Search for Planets (SuperWASP; \citealt{2010A&A...520L..10B}), and VarAstro\footnote{\url{https://var.astro.cz/en/}} (previously known as the O–C Gateway). Times of minima from AAVSO, SuperWASP, and TESS short cadence data were obtained directly by fitting a parabola to the light curve near the eclipse minimum. ASAS-SN, ZTF (\citealt{2019PASP..131f8003B,2019PASP..131a8003M}), and TESS 1800s-cadence light curves were first phase-folded with the period-shift technique of \cite{2020AJ....159..189L} before calculating the eclipse timings. All Heliocentric Julian Dates (HJD) were then converted to Barycentric Julian Date (BJD) in Barycentric Dynamical Time using the online tool developed by \cite{2010PASP..122..935E}. Data from VarAstro that lack reported errors were assigned uniform uncertainties of $0.001$ for CCD observations. The eclipse timings derived from the photometric data acquired in this study are provided in Table \ref{Tab:extracted-mins}. The complete collection of minima timings for the studied contact binaries has been compiled in a machine-readable format and made available online.

\begin{table*}
\caption{Times of minima measured from the ground-based observational data.}
\centering
\small
\begin{tabular}{c c c c c}
\hline
System & Min.($BJD_{TDB}$) & Error  & Epoch  & O-C\\ 
\hline
G1721  &2460795.4190	     &0.0003	&-0.5	   &-0.0005  \\
       &2460795.5666 	     &0.0003	&0	     &0     \\
G2161  &2460796.3742 	     &0.0006	&0	     &0     \\
       &2460796.5212 	     &0.0006	&0.5	   &-0.0021  \\
NO Leo &2460779.5268 	     &0.0001	&0	     &0     \\
       &2460779.6708 	     &0.0001	&0.5	   &-0.0015  \\
TIC422347573     &2460797.4300 	&0.0016 	&0   &0   \\
                 &2460797.5601 	&0.0002 	&0.5 	&0.0037 \\ 
                 &2460822.4474 	&0.0003 	&99 	&0.0008 \\               
V337 UMa         &2460713.3648 	&0.0006 	&-102	&-0.0020      \\
                 &2460713.5571 	&0.0008 	&-101.5	&-0.0013    \\
                 &2460752.4697 	&0.0005 	&0	    &0          \\
                 &2460752.6549 	&0.0010 	&0.5	  &-0.0065    \\
V369 Boo         &2460771.4869 	&0.0001 	&-51	&0.0000       \\
                 &2460787.4076 	&0.0003 	&-3	  &0.0002       \\
                 &2460787.5759 	&0.0023 	&-2.5	&0.0027       \\
                 &2460788.4024 	&0.0004 	&0	  &0            \\
                 &2460788.5687 	&0.0003 	&0.5	&0.0004       \\
V475 Ser         &2460812.5075 	&0.0015 	&0      	&0   	  \\    
                 &2460813.4687 	&0.0054 	&4        &0.0016 	  \\
                 &2460813.5887 	&0.0007 	&4.5    	&0.0016 	  \\
V640 Aur         &2460674.3980 	&0.0002 	&0	  &  0          \\
                 &2460674.5627 	&0.0002 	&0.5	&  0.0006     \\
                 &2460674.7260 	&0.0003 	&1	  & -0.0001     \\  
V689 Vir         &2460793.6189 	&0.0001 	&-0.5 	&-0.0002  \\
                 &2460793.7590 	&0.0002 	&0   	    &0        \\
V702 Aur         &2460672.3934 	&0.0006 	&-3.5 	&-0.0044  \\   
                 &2460672.5612 	&0.0007 	&-3 	&-0.0036  \\
                 &2460673.3950 	&0.0006 	&-0.5 	&-0.0051  \\
                 &2460673.5672 	&0.0008 	&0	      &0        \\
\hline
\end{tabular}
\label{Tab:extracted-mins}
\end{table*}

\begin{figure*}
\centering
\includegraphics[height=0.95\textheight]{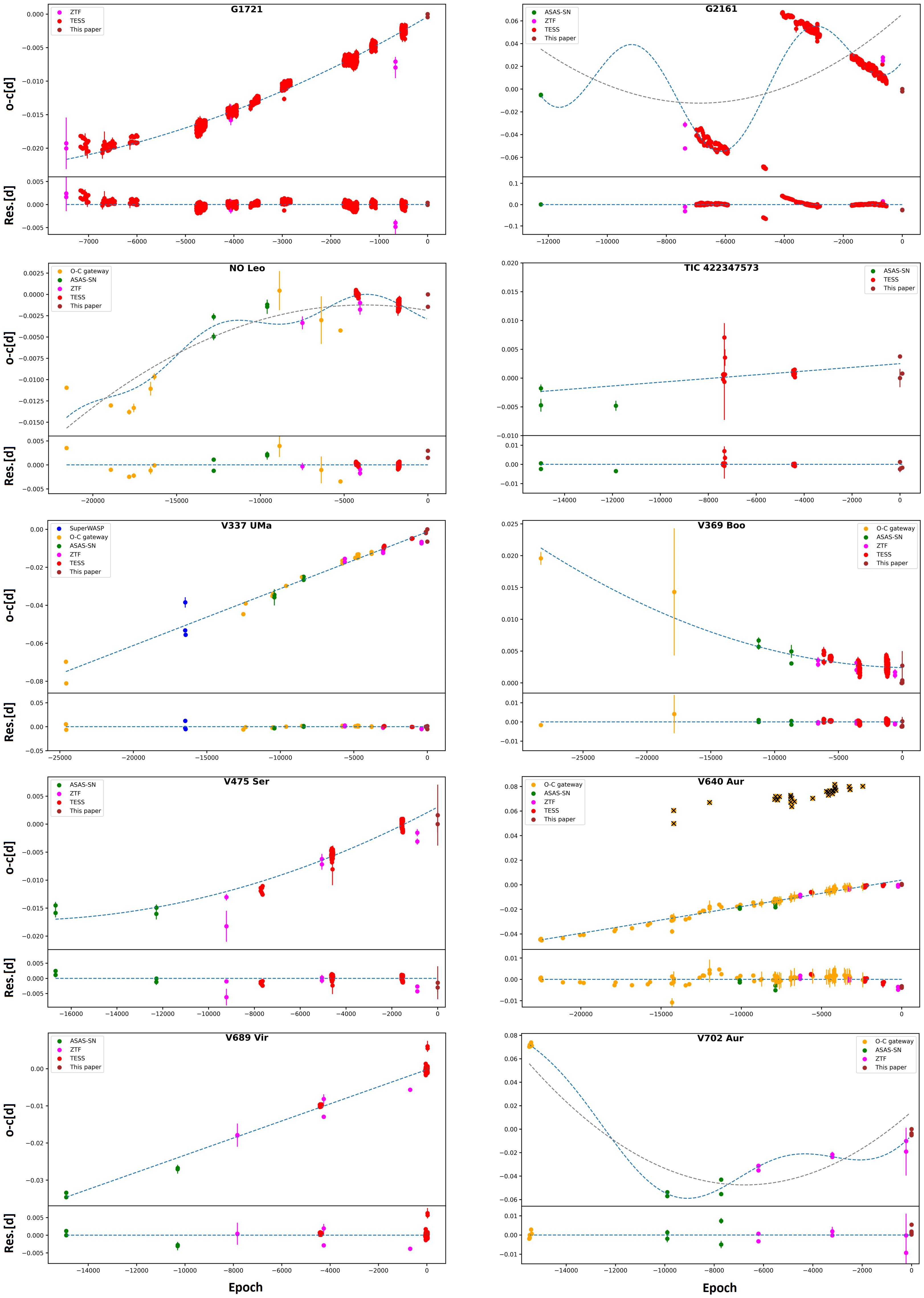}
\caption{O-C diagrams of the target systems. Colored points show eclipse timings from this work or the literature. The appropriate linear, quadratic, or cyclic fit for each system is overplotted to illustrate period variations.}
\label{fig:all_oc}
\end{figure*}

O-C values were computed with respect to the following equation:
\begin{equation}
\mathrm{BJD} = \mathrm{BJD}_{0} + P \times E,
\end{equation}
where BJD is the observational eclipse timings, $\mathrm{BJD}_{0}$ is the initial epoch, $P$ is the orbital period (the two parameters can be found in Table \ref{Tab:ephemeris}), and $E$ is the cycle number. The calculated O–C values are listed in Table \ref{Tab:extracted-mins}, along with a machine-readable version of the data. Figure \ref{fig:all_oc} illustrates the O–C diagrams derived for all target binaries, where six targets exhibit long-term variations. Only a linear correction was applied for TIC 422347573, V337 UMa (the period of V337 UMa was obtained through a pre-linear correction), V640 Aur, and V689 Vir. The following equation was used to fit the O-C diagrams for G1721, V369 Boo, and V475 Ser:

\begin{equation}
\mathrm{O\!-\!C} = \Delta T_{0} + \Delta P_{0}\times E + \frac{\beta}{2}\times E^{2}.
\end{equation}

Besides the secular trend, cyclic variations are also present in the O–C diagrams of V702 Aur, NO Leo, and G2161. The following equation was used to describe these variations:

\begin{equation}
	\mathrm{O\!-\!C} = \Delta T_0 + \Delta P_0 \times E 
	+ \frac{\beta}{2} \times E^2 
	+ A \times \sin\!\left(\frac{2\pi}{P_3} E + \phi \right).
\end{equation}

The observed cyclic behavior may originate from either magnetic activity within the binary components or the light travel time effect induced by an additional companion; These two possibilities are examined in detail in Section 6. The results of the O-C fitting are shown in Table \ref{tab:oc_coeff}. It is worth noting that data points for V640 Aur that clearly deviate from the fitting curve were excluded, and a linear fit was performed. These excluded points originated from the VarAstro database and have very similar reported timing uncertainties, with a mean value of approximately 0.004 d. Moreover, for the other targets, some outlier points are visible in the O–C curves. According to our analysis, these may result from insufficient measurement accuracy or magnetic activity in the targets. Outliers were intentionally kept in the analysis to preserve the original dataset.

The results indicate that five targets show an upward trend, implying a long-term increase in their orbital periods; one binary exhibit a downward trend, pointing to a long-term decade in the orbital periods; and the periods of the remaining four systems appear essentially constant. The updated linear ephemerides are provided in Table \ref{Tab:ephemeris}.

\renewcommand\arraystretch{1.2}
\begin{table*}
	\centering
	\caption{The values of initial epoch, the initial $P$, the corrected initial epoch, and the corrected orbital period for the ten targets in O-C analysis.}
\label{Tab:ephemeris}
\begin{tabular}{lcccc}
\hline
Target & $T_0$ (BJD) & $P$(d) & Corrected $T_0$ (BJD) & Corrected $P$(d)\\
\hline
G1721         & 2460795.5666     &0.294290     &2460795.5662 (1)    &0.294294 (1) \\
G2161         & 2460796.3742     &0.298145     &2460796.4401 (10)   &0.298168 (1)  \\
NO Leo        & 2460779.5268     &0.290910     &2460779.5249 (4)    &0.290910 (1)  \\
TIC 422347573 & 2460797.4300     &0.252694     &2460797.4325 (7)    &0.252694 (1) \\
V337 UMa      & 2460752.4697     &0.383362     &2460752.4683 (5)    &0.383365 (1) \\
V369 Boo      & 2460788.4024     &0.331677     &2460788.4048 (1)    &0.331677 (1) \\
V475 Ser      & 2460812.5075     &0.239904     &2460812.5105 (2)    &0.239906 (1)\\
V640 Aur      & 2460674.3980     &0.327996     &2460674.4019 (1)    &0.327998 (1) \\
V689 Vir      & 2460793.7590     &0.279968     &2460793.7587 (1)    &0.279970 (1) \\
V702 Aur      & 2460673.5672     &0.334130     &2460673.5821 (23)   &0.334148 (1)  \\
\hline
\end{tabular}
\end{table*}

\begin{table*}
\centering
\scriptsize
\caption{Fitting coefficients and associated uncertainties derived from the O-C analysis of the target systems.}
\label{tab:oc_coeff}
\begin{tabular}{lcccccccccccccc}
\hline
Target & $\Delta T_{0}$ & Error & $\Delta P_{0}$ & Error &
$\beta$ & Error & $A_3$ & Error & $P_{3}$ & Error & $\phi$ & Error & d$M$/d$t$ & Error\\[2pt]
&
\multicolumn{2}{c}{($10^{-4}$\,d)} &
\multicolumn{2}{c}{($10^{-7}$\,d)} &
\multicolumn{2}{c}{($10^{-7}$\,d\,yr$^{-1}$)} &
\multicolumn{2}{c}{(d)} &
\multicolumn{2}{c}{(yr)} &
\multicolumn{2}{c}{(rad)} &
\multicolumn{2}{c}{($10^{-7}\,M_{\odot}\,\text{yr}^{-1}$)}\\
\hline
G1721          &-3.89   &0.44     &42.67  &0.32 &4.70    &0.13 &         &         &     &     &      &     &-3.43  &0.09      \\
G2161          &659.09  &9.70     &227.40 &6.68 &40.45   &2.21 &0.04346  &0.00052  &4.79 &0.04 &-1.29 &0.03 &7.04   &0.38      \\
NO Leo         &-18.96  &3.62     &-3.45  &0.97 &-1.15   &0.11 &-0.00125 &0.00018  &7.10 &0.29 &0.98  &0.20 &0.64   &0.06      \\
TIC 422347573  &24.98   &6.82     &3.23   &1.00 &        &     &         &         &     &     &      &     &       &      \\
V337 UMa       &-13.40  &4.93     &29.93  &0.64 &        &     &         &         &     &     &      &     &       &      \\
V369 Boo       &23.99   &0.71     &-0.47  &0.32 &0.48    &0.04 &         &         &     &     &      &     &2.80   &0.23      \\
V475 Ser       &30.37   &1.66     &21.75  &0.65 &1.78    &0.15 &         &         &     &     &      &     &-1.20  &0.10      \\
V640 Aur       &38.56   &1.09     &21.67  &0.17 &        &     &         &         &     &     &      &     &       &      \\
V689 Vir       &-2.67   &0.65     &23.02  &0.25 &        &     &         &         &     &     &      &     &       &      \\
V702 Aur       &148.92  &23.36    &184.32 &6.42 &29.73   &1.04 &0.02091  &0.00262  & 8.61&0.75 &-14.38&0.31 &-126.76&4.43      \\
\hline
\end{tabular}
\end{table*}

\vspace{0.6cm}
\section{Light Curve Solutions}
The light curves were modeled with the BSN v.1.0 software package (\citealt{2025Galax..13...74P}), a dedicated application developed for the analysis of contact binary stars. A detailed description of its architecture and algorithms is provided by \cite{2025Galax..13...74P}.

As listed in Table \ref{Tab:tess}, with the exception of V702 Aur, ground-based photometric data were utilized in conjunction with TESS data. To minimize the effect of spot evolution and activity-related light curve changes, we selected one of the most recent TESS sectors close in time to our ground-based observations (Table \ref{Tab:tess}). However, the selected data were checked against other recent sectors to ensure no significant differences. An exception was TIC 422347573, for which the available TESS time-series data in both sectors had exposure times of 600\,s and 1800\,s. As noted by \cite{kipping2010binning}, relatively long exposure times compared to the orbital period can introduce a finite integration-time effect, in which the observed flux is averaged over the exposure duration. This effect acts as a smoothing filter on the light curve and can produce shallower and more rounded eclipse minima. Therefore, to avoid possible distortions in the light curve morphology caused by the long TESS exposure times, we primarily used the ground-based photometric data for the analysis of TIC 422347573. The resulting model obtained from the ground-based $R_c$ data was subsequently compared with the TESS observations from both sectors after accounting for the finite integration-time effect described by \cite{kipping2010binning}, and was found to reproduce the observed light curves satisfactorily.

A consistent set of physical and modeling assumptions was established at the outset of the photometric analysis of the target systems. All targets were modeled under the assumption of a contact configuration, as their light curve morphology, catalog classifications, and short orbital periods are all consistent with stars sharing a common envelope and remaining in thermal contact. Orbital phases were determined using the ephemerides listed in Section 3. Gravity-darkening coefficients were fixed at $g_{1}=g_{2}=0.32$ (\citealt{1967ZA.....65...89L}), and bolometric albedos were adopted as $A_{1}=A_{2}=0.5$ (\citealt{1969AcA....19..245R}). The radiative properties of each component are described using the atmospheric model of \cite{2004AA...419..725C}. The BSN application employs the logarithmic limb-darkening laws with coefficients adopted from \cite{1993AJ....106.2096V} and the updated tabulations provided on the official webpage\footnote{\url{https://faculty.fiu.edu/~vanhamme/limb-darkening/}}, including those for the TESS passband. Effective temperatures ($T$) for the target systems were first adopted from the Gaia DR3 catalog (Table \ref{Tab:systemsinfo}). For V640 Aur, which lacks a Gaia DR3 entry, had its temperature adopted from version 8.2 of the TIC. Based on the relative depths of the light curve minima, the initial temperatures listed in Gaia DR3 and the TIC were assumed to represent the hotter component of each binary system. The observed contrast between the primary and secondary eclipse depths provided the basis for determining the effective temperature of the cooler component.

\begin{figure*}
\centering
\includegraphics[width=0.99\textwidth]{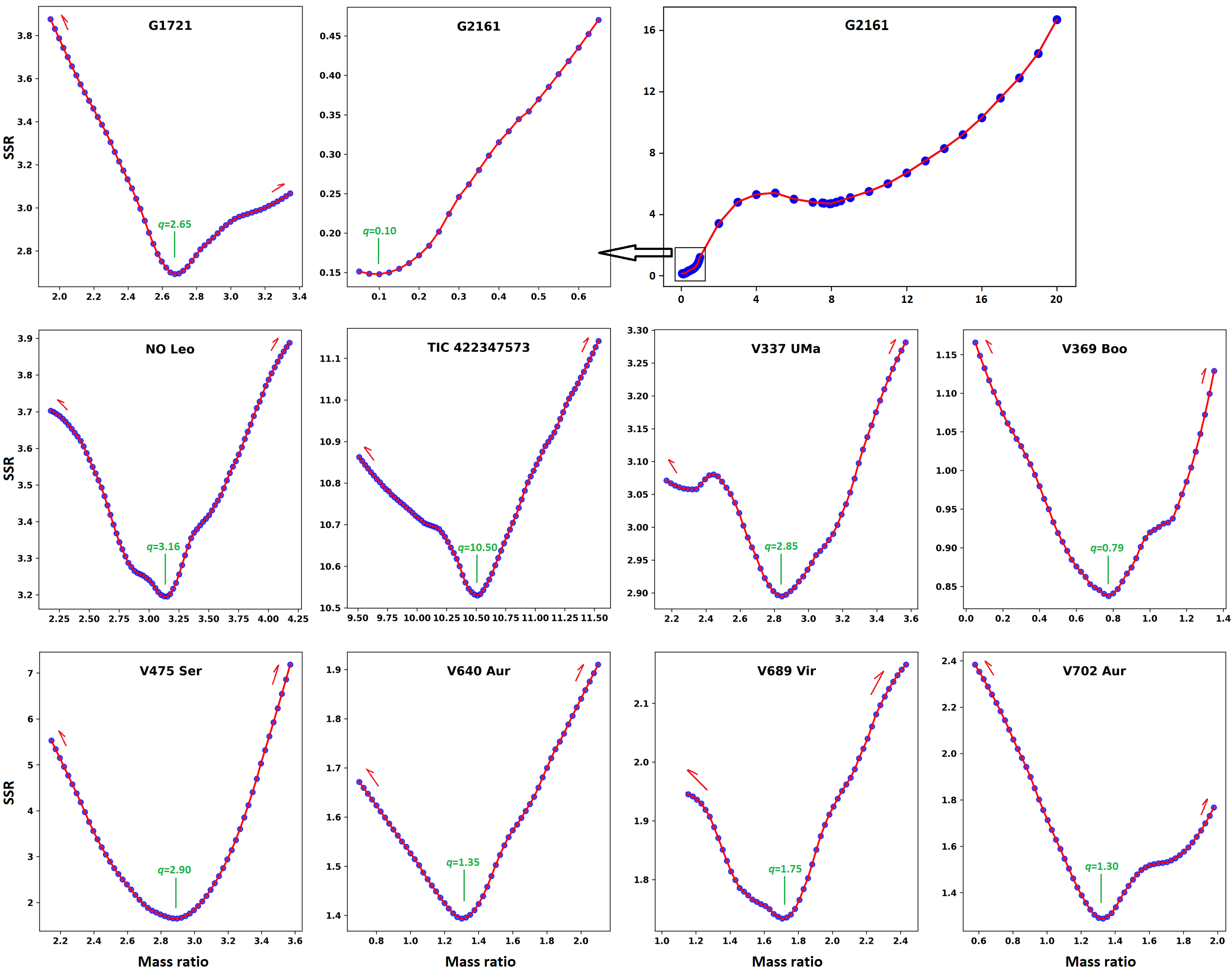}
\caption{$q$-search diagrams for the 10 analyzed systems. The sum of squared residuals as a function of the mass ratio is shown for each system, where $\mathrm{SSR} = \sum (O - C)^2$, and the minimum of each curve shows the optimal photometric $q$. The full search range is shown for G2161, which has the lowest mass ratio among the analyzed systems, along with a zoomed-in view around the minimum. The red arrow indicates the general trend, while the green line marks the searched mass ratio value.}
\label{Fig:q}
\end{figure*}

\begin{figure*}
\centering
\includegraphics[width=0.99\textwidth]{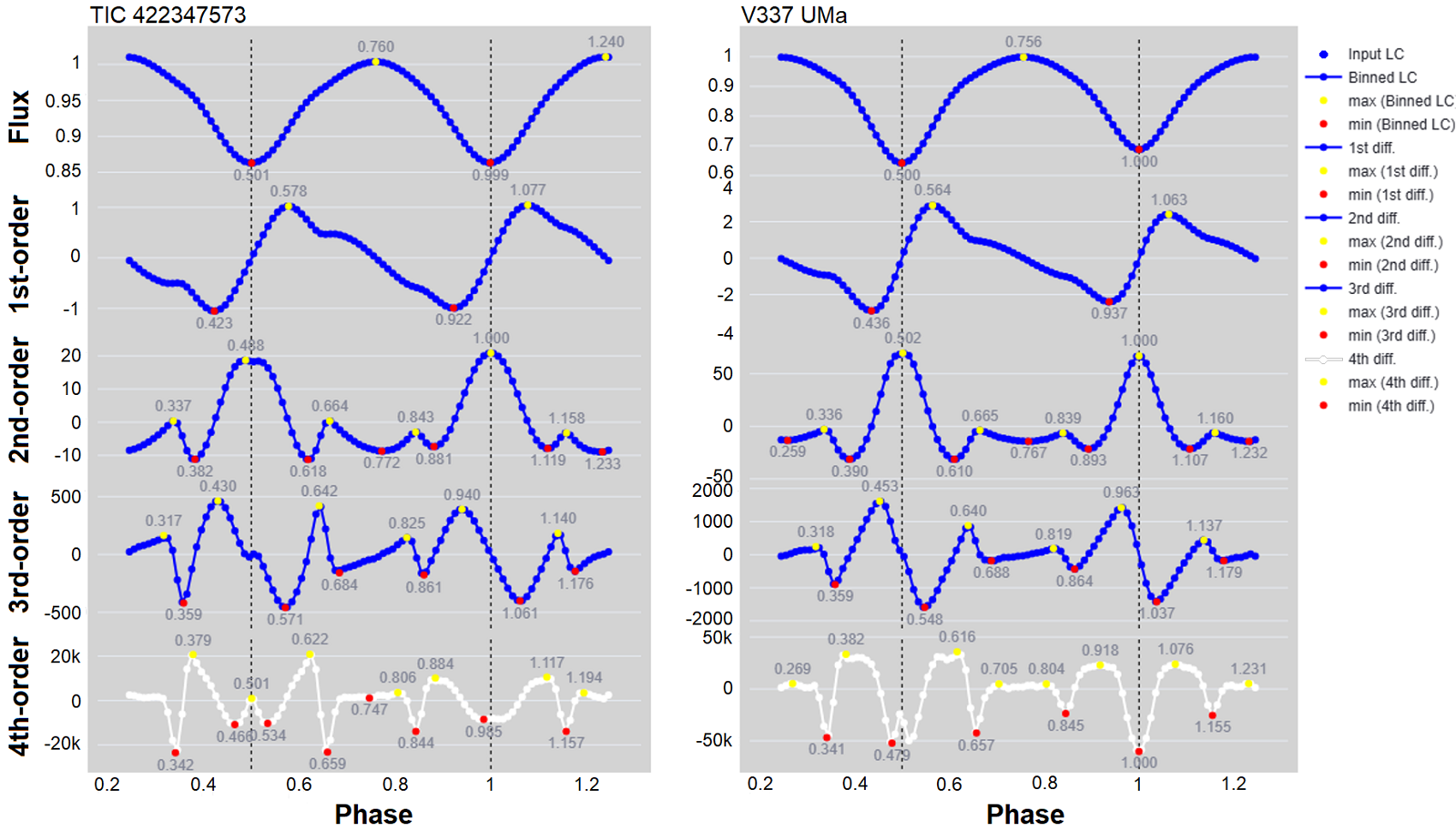}
\caption{Observed photometric light curves and their first through fourth time derivatives (ordered from top to bottom) for systems TIC 422347573 (left) and V337 UMa (right), shown here as illustrative examples, illustrating the process of estimating the initial mass ratio for each target system using the Kouzuma method. The vertical-axis units correspond to W~m$^{-2}$, 10~W~m$^{-2}$~day$^{-1}$, $10^{2}$~W~m$^{-2}$~day$^{-2}$, $10^{4}$~W~m$^{-2}$~day$^{-3}$, and $10^{6}$~W~m$^{-2}$~day$^{-4}$, respectively.}
\label{Fig:q-K}
\end{figure*}

Two approaches were employed to obtain initial estimates of the mass ratio ($q$) for the purpose of starting the light curve analysis. First, the $q$-search method was applied, following the approach of \cite{2005ApSS.296..221T}. The mass ratio parameter was first sampled over the interval 0.05-20 with increments of 0.1. After identifying the approximate location of the minima, a narrower interval around each minimum was re-examined with a finer step size of 0.025 to refine the estimates. The procedure was implemented through a Python script using an automated loop to systematically compute models over the defined grid. For each trial value of $q$, the sum of squared residuals between the synthetic and observed light curves was calculated, and the global minimum was adopted as the optimal photometric mass ratio (Figure~\ref{Fig:q}). Second, a technique for estimating the photometric mass ratio of overcontact binaries was presented by \cite{2023ApJ...958...84K, 2025PASJ..tmp..108K}, relying on the analysis of higher-order light-curve derivatives. Rather than relying on iterative optimization procedures, this approach infers the mass ratio from diagnostic features in the second- and third-order light curve derivatives. Application of the method involves computing the third-order derivative, identifying local extrema near the eclipse phases, and combining these with the orbital period to derive a parameter $W$, which exhibits a strong correlation with $q$. When evaluated against spectroscopic mass-ratio measurements, photometric estimates agree within the reported uncertainties for about 67\% of the sample, whereas nearly 95\% remain within an absolute difference of $\pm0.1$. The method requires clear maxima and minima in the relevant derivatives, as discussed in \cite{2023ApJ...958...84K, 2025PASJ..tmp..108K}. Results obtained using the two approaches are summarized in Table \ref{Tab:lc-analysis}. The mass ratio estimation procedure is illustrated in Figure \ref{Fig:q-K} for two representative target systems, one characterized by a low mass ratio and the other by a higher mass ratio. It should be noted that, in accordance with the Kouzuma method code, a bin number of 100 was adopted in our estimation (\citealt{2023ApJ...958...84K}). We applied this derivative-based approach to the target systems in the present study. It should be noted that the $q$-search using the first method was performed with at least two of the available datasets in the TESS, $V$, and $R_c$ bands for each target. The second method relied exclusively on TESS data, except for V702 Aur, for which $V$-band data were used. The initial mass ratio was taken as the average of the estimates from both methods, which were in close agreement. Iterative light curve modeling was initialized with this estimated value, leading to the final mass ratio solutions for the systems.

An asymmetry between the two light curve maxima, commonly referred to as the O'Connell effect, was detected in five of the targets. To achieve satisfactory fits to the observed light curves, a cool starspot was introduced on one of the binary components (Table \ref{Tab:lc-analysis}). Each system was examined with a cool or hot starspot placed on either component at the light curve maxima. The configuration yielding the smallest $\chi^2$ was selected, unambiguously favoring the chosen location and temperature of the starspot in all seven systems. This phenomenon is generally attributed to magnetic activity on the stellar surface and is explained by the presence of starspots, although other physical mechanisms have also been suggested in the literature (e.g., \citealt{1990ApJ...355..271Z}; \citealt{2003ChJAA...3..142L}).

We utilized the ground- and space-based photometric data along with the initial parameter estimates to construct a theoretical fit of the light curves. The BSN application's optimization tool was then applied to refine the solutions, resulting in more tightly constrained values for the main parameters, including component effective temperatures ($T_{1,2}$), mass ratio ($q$), fillout factor ($f$), and orbital inclination ($i$). The BSN application employs a generation-based evolutionary algorithm to optimize the parameters, iteratively improving candidate solutions over successive generations (\citealt{back2023evolutionary}). The final parameter solutions and their uncertainties were obtained using the Markov Chain Monte Carlo (MCMC) method. The BSN application provides efficient computation for MCMC fitting, allowing rapid generation of synthetic light curves. In the simulations, we employed 26 walkers and ran 3000 iterations to sample the five principal parameters ($T_1$, $T_2$, $q$, $f$, and $i$). The burn-in phase consisted of the first 500 iterations of each walker, which were removed before the analysis to ensure convergence. Posterior distributions from the remaining samples were then used to determine both the best-fitting values and the corresponding 1$\sigma$ uncertainties for each parameter, providing a robust estimate of the model's reliability. Therefore, the median of the posterior distribution was adopted as the final reported value (Table \ref{Tab:lc-analysis}) for all parameters derived from the MCMC analysis in the BSN application, providing a robust central estimate even for asymmetric distributions.

\begin{figure*}
\centering
\includegraphics[width=0.7\textwidth]{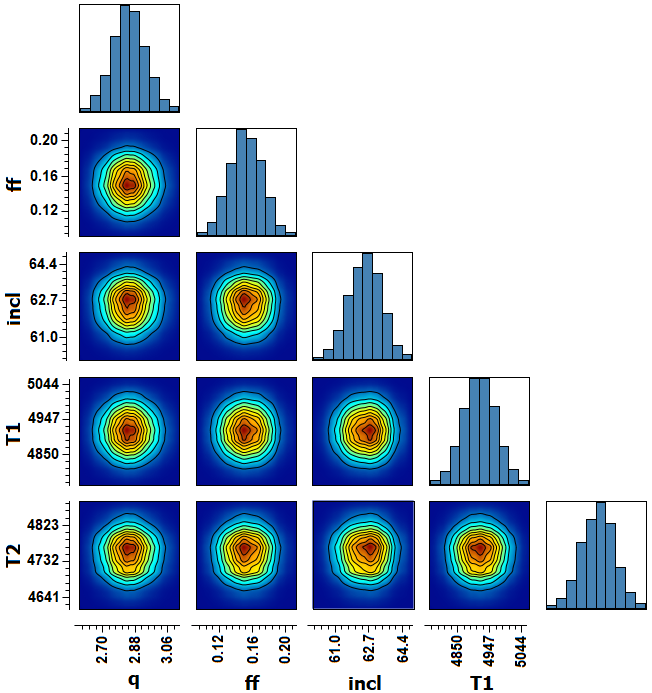}
\caption{Corner plot of the MCMC results for V475 Ser, presented as a representative case from the analyzed sample. The distributions of the posterior samples and the pairwise correlations among the five fitted parameters are displayed.}
\label{Fig:corner}
\end{figure*}

\begin{figure*}
\centering
\includegraphics[width=1\textwidth]{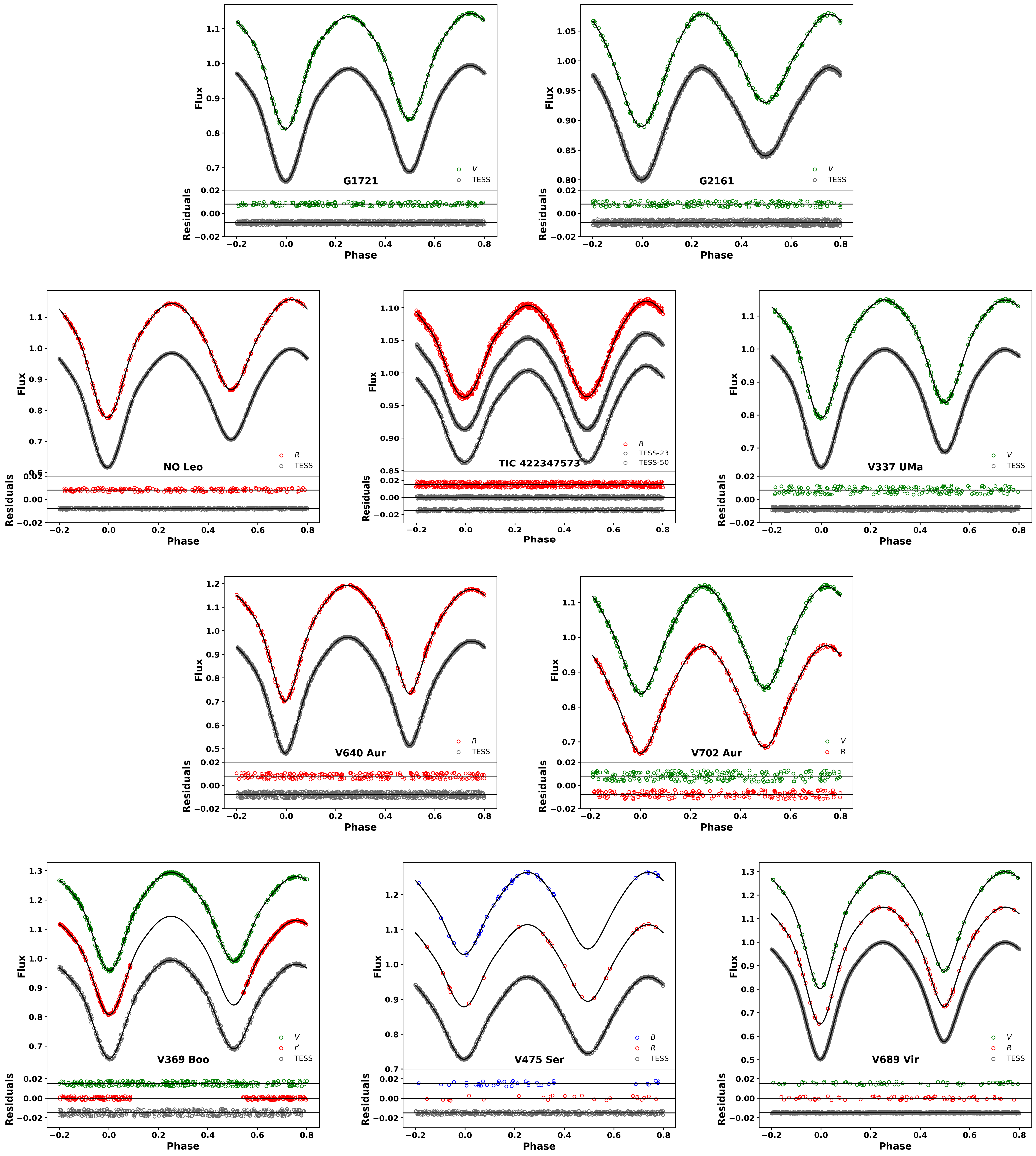}
\caption{Best-fitting synthetic models overlaid on the observed light curves of the 10 contact binary systems. Different colors denote the photometric filters, and the model residuals are presented in the lower panels.}
\label{Fig:lc}
\end{figure*}

\begin{figure}
\centering
\includegraphics[width=0.4\textwidth]{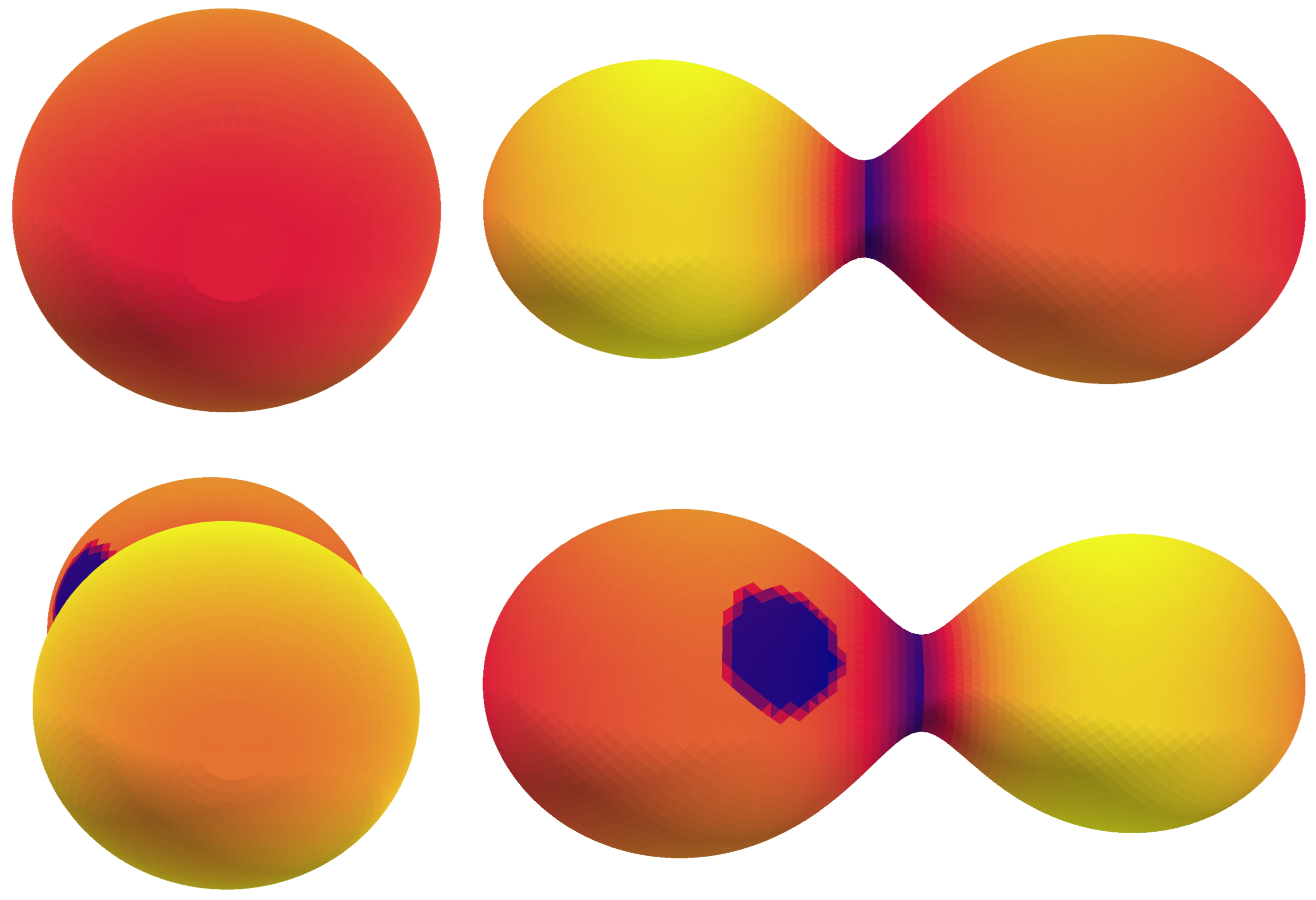}
\caption{Three-dimensional views of the V640 Aur binary system at orbital phases 0.00, 0.25, 0.50, and 0.75, shown as a representative example of the analyzed contact binaries to illustrate the orbital configuration of the stellar components together with the adopted starspot.}
\label{Fig:3d}
\end{figure}

Corner plots illustrating the posterior distributions and parameter correlations from the MCMC analysis are presented for the V475 Ser system as an example in Figure \ref{Fig:corner}. The best-fit parameters derived from the light curve analysis, along with their associated uncertainties, are provided in Table \ref{Tab:lc-analysis}. Figure \ref{Fig:lc} shows the final synthetic light curves overlaid on the observed photometric data for all target binaries. In addition, three-dimensional (3D) models were generated for all modeled systems, and the result for V640 Aur is presented in Figure \ref{Fig:3d} as a representative example.

\begin{sidewaystable*}
\renewcommand\arraystretch{1.5}
\setlength{\tabcolsep}{4pt}
\caption{Results of the light curve modeling for the 10 analyzed binary systems.}
\centering
\footnotesize
\begin{tabular}{c c c c c c c c c c c}
\hline
Parameter & G1721 & G2161 & NO Leo & TIC 422347573 & V337 UMa & V369 Boo & V475 Ser & V640 Aur & V689 Vir & V702 Aur\\
\hline
$T_{1}$ (K) 	&	$5844_{\rm-(41)}^{+(38)}$ 	&	$5825_{\rm-(49)}^{+(50)}$ 	&	$4911_{\rm-(40)}^{+(40)}$ 	&	$5919_{\rm-(50)}^{+(45)}$ 	&	$5604_{\rm-(45)}^{+(43)}$ 	&	$5922_{\rm-(27)}^{+(26)}$ 	&	$4912_{\rm-(49)}^{+(49)}$ 	&	$5609_{\rm-(46)}^{+(46)}$ 	&	$5233_{\rm-(35)}^{+(38)}$ 	&	$5983_{\rm-(44)}^{+(45)}$\\
$T_{2}$ (K) 	&	$5604_{\rm-(43)}^{+(36)}$ 	&	$5102_{\rm-(45)}^{+(47)}$ 	&	$4451_{\rm-(33)}^{+(32)}$ 	&	$5944_{\rm-(51)}^{+(53)}$ 	&	$5278_{\rm-(40)}^{+(38)}$ 	&	$5535_{\rm-(24)}^{+(24)}$ 	&	$4756_{\rm-(48)}^{+(43)}$ 	&	$5388_{\rm-(46)}^{+(46)}$ 	&	$4892_{\rm-(39)}^{+(42)}$ 	&	$5827_{\rm-(45)}^{+(47)}$\\
$q=M_2/M_1$ 	&	$2.647_{\rm-(103)}^{+(104)}$ 	&	$0.127_{\rm-(13)}^{+(11)}$ 	&	$3.177_{\rm-(146)}^{+(90)}$ 	&	$10.510_{\rm-(26)}^{+(22)}$ 	&	$2.869_{\rm-(192)}^{+(117)}$ 	&	$0.831_{\rm-(26)}^{+(12)}$ 	&	$2.846_{\rm-(91)}^{+(94)}$ 	&	$1.400_{\rm-(49)}^{+(48)}$ 	&	$1.731_{\rm-(70)}^{+(78)}$ 	&	$1.278_{\rm-(86)}^{+(82)}$\\
$i^{\circ}$ 	&	$72.08_{\rm-(29)}^{+(28)}$ 	&	$64.10_{\rm-(94)}^{+(1.13)}$ 	&	$71.19_{\rm-(44)}^{+(56)}$ 	&	$62.39_{\rm-(88)}^{+(92)}$ 	&	$73.54_{\rm-(69)}^{+(46)}$ 	&	$69.97_{\rm-(13)}^{+(14)}$ 	&	$62.76_{\rm-(89)}^{+(79)}$ 	&	$81.40_{\rm-(93)}^{+(92)}$ 	&	$80.14_{\rm-(70)}^{+(76)}$ 	&	$66.05_{\rm-(69)}^{+(66)}$\\
$f$ 	&	$0.038_{\rm-(8)}^{+(8)}$ 	&	$0.386_{\rm-(45)}^{+(49)}$ 	&	$0.141_{\rm-(17)}^{+(18)}$ 	&	$0.051_{\rm-(17)}^{+(19)}$ 	&	$0.024_{\rm-(3)}^{+(5)}$ 	&	$0.038_{\rm-(4)}^{+(5)}$ 	&	$0.152_{\rm-(20)}^{+(20)}$ 	&	$0.197_{\rm-(38)}^{+(36)}$ 	&	$0.119_{\rm-(28)}^{+(26)}$ 	&	$0.197_{\rm-(33)}^{+(41)}$\\
$\Omega_1=\Omega_2$ 	&	6.12(4)	&	2.01(4)	&	6.76(9)	&	15.64(7)	&	6.43(4) 	&	3.45(2)	&	6.32(9)	&	4.26(12)	&	4.80(7)	&	4.08(11)\\
$l_1/l_{tot}$ 	&	0.323(5)	&	0.908(7)	&	0.357(4)	&	0.107(4)	&	0.323(3) 	&	0.601(4)	&	0.310(4) 	&	0.462(8)	&	0.446(6)	&	0.476(7)\\
$l_2/l_{tot}$ 	&	0.677(4)	&	0.092(2)	&	0.643(4)	&	0.893(4)	&	0.677(4) 	&	0.399(4)	&	0.690(4) 	&	0.538(8)	&	0.554(6)	&	0.524(7)\\
$l_3/l_{tot}$ 	&	-	&	-	&	-	&	-	&	-	&	-	&	-	&	-	&	-	&	-\\
$r_{(mean)1}$ 	&	0.301(3)	&	0.574(10)	&	0.293(10)	&	0.204(2)	&	0.293(1) 	&	0.400(3)	&	0.302(10)	&	0.368(16)	&	0.342(9)	&	0.376(15)\\
$r_{(mean)2}$ 	&	0.470(3)	&	0.236(11)	&	0.491(8)	&	0.586(2)	&	0.476(2) 	&	0.367(3)	&	0.483(10)	&	0.426(15)	&	0.438(9)	&	0.419(15)\\
\hline																				
$Col.^\circ$(spot) 	&	98(1)	&	- 	&	114(2) 	&	114(1)	&	- 	&	107(1)	&	- 	&	107(1) 	&	- 	&	-\\
$Long.^\circ$(spot) 	&	74(1)	&	- 	&	119(2) 	&	69(1)	&	- 	&	319(3)	&	- 	&	305(2) 	&	- 	&	-\\
$Radius^\circ$(spot) 	&	12(1)	&	- 	&	17(1) 	&	11(1)	&	- 	&	21(1)	&	- 	&	18(1) 	&	- 	&	-\\
$T_{spot}/T_{star}$ 	&	0.91(1)	&	- 	&	0.94(1) 	&	0.94(1)	&	- 	&	0.88(1)	&	- 	&	0.89(1) 	&	- 	&	-\\
Component 	&	Secondary 	&	- 	&	Secondary 	&	Secondary 	&	- 	&	Secondary	&	- 	&	Secondary 	&	- 	&	-\\
\hline
Kouzuma Method* & 0.35(4) & 0.15(4) & 0.30(4) & 0.13(3) & 0.33(4) & 0.88(4) & 0.32(4) & 0.75(4) & 0.51(4) & 0.76(3) \\
$q$-search Method & 2.65 & 0.10 & 3.16 & 10.50 & 2.85 & 0.79 & 2.90 & 1.35 & 1.75 & 1.30 \\
\hline
\end{tabular}
\label{Tab:lc-analysis}
\footnotesize \textit{*} In the Kouzuma method, the more massive component is consistently designated as $M_1$, and the less massive component as $M_2$.
\end{sidewaystable*}

\vspace{0.6cm}
\section{Estimation Absolute Parameters}
Absolute stellar parameters for W UMa binary stars can be calculated through various ways, with empirical parameter relationships being among the most widely used methods. Studies of orbital period variations in contact binaries have consistently shown a strong connection between the orbital period and the systems' physical and evolutionary properties. Previous studies by \cite{2001MNRAS.328..635Q,2001MNRAS.328..914Q} demonstrated that changes in the orbital period are correlated with the mass of the more massive component ($M_1$). In addition, \cite{2003MNRAS.342.1260Q} identified two narrower, parallel sequences in the $P$–$M_{1}$ distribution for overcontact binaries. Subsequent works refined these relationships further: \cite{2006MNRAS.370L..29G} demonstrated that primary masses increase systematically with period, while secondary masses remain nearly independent of it; \cite{2006MNRAS.373.1483E} derived a logarithmic $P$-$M_{\rm total}$ relation for W UMa binaries; and \cite{2008MNRAS.390.1577G} showed that the orbital period alone can estimate component masses with an accuracy of about 15\%. Subsequent analyses focused on period-defined groups of contact binaries. For instance, \cite{2015AJ....150...69Y} found that a decreasing $P$ in deep, low $q$ contact stars corresponds to total mass loss, while \cite{2018PASJ...70...90K} derived a $P$–$M_{1}$ relation for short-period Kepler binaries. Additionally, \cite{2021ApJS..254...10L} provided updated period–mass relationships for both components of contact binaries. Then, \cite{2022MNRAS.510.5315P} conducted a comprehensive re-evaluation of the period–mass correlations, utilizing a sample of 118 systems that combined newly estimated masses derived from the Gaia DR3 parallax method with data from the literature. The study \cite{2023A&A...672A.176P} also discusses mass variations in a large sample of contact binary stars and their possible correlation with orbital period.

\begin{figure*}
\centering
\includegraphics[width=0.99\textwidth]{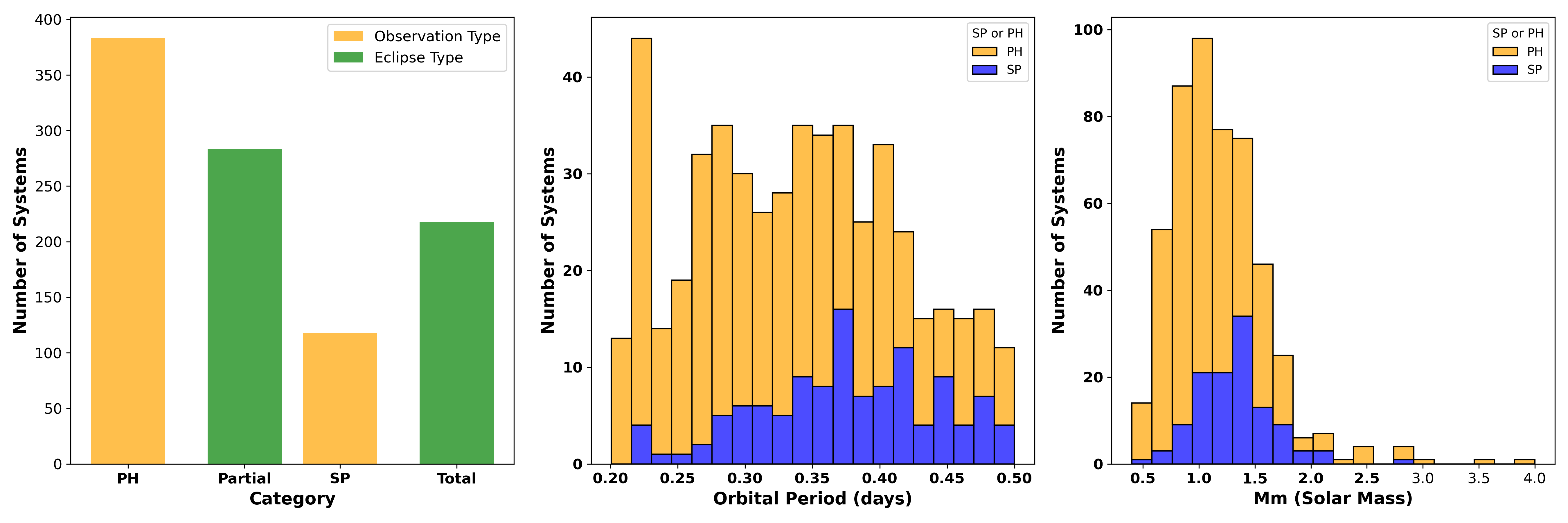}
\caption{Summary statistics are shown for the sample of used 502 systems. The bar and histogram plots illustrate the distribution of observation type, eclipse type, orbital period, and mass. The sample contains 381 PH and 121 SP systems, 283 partial and 219 total eclipses, with orbital periods ranging from 0.2 to 0.5 days and primary component masses from 0.4 to 4~$M_\odot$.}
\label{Fig:P-Mm-sample}
\end{figure*}
\begin{figure*}
\centering
\includegraphics[width=0.99\textwidth]{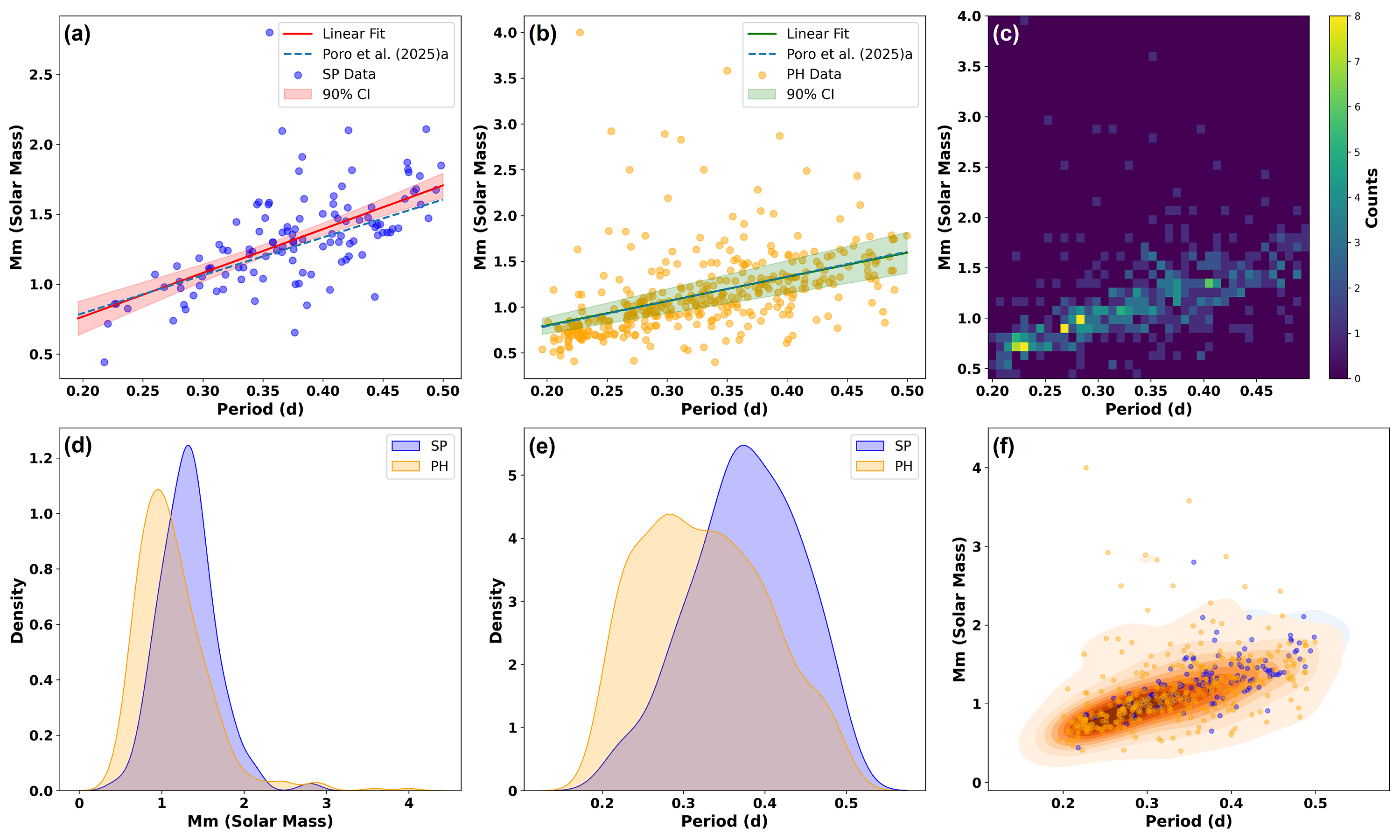}
\caption{(a) Linear fit SP, (b) Linear fit PH, (c) 2D histogram heatmap, (d) KDE of $M_m$, (e) KDE of orbital period, (f) 2D KDE of orbital period versus $M_m$.}
\label{Fig:P-Mm}
\end{figure*}

The empirical relationship between orbital period and the mass of the primary component is revisited in this work. The empirical estimation of the mass of the more massive component ($M_{\rm m}$) was carried out using a sample of 502 systems from \cite{2025MNRAS.538.1427P}, restricted to systems with reported masses for both components and with limited orbital periods. The sample includes W UMa contact systems with orbital periods shorter than 0.5 days and for which masses have been reported for both components; the application of these two criteria reduced the original sample of 818 systems accordingly. The $P$–$M_{1}$ relation was also examined in the study of \cite{2025MNRAS.538.1427P}; however, in their analysis, $M_{1}$ was adopted as reported in the literature, without considering the stellar subtype, whereas we assumed that the more massive component corresponds to $M_{1}$. Furthermore, unlike their approach, in which an upper temperature limit of 7000 K was imposed based on \cite{2021ApJS..254...10L}, we did not apply any temperature constraint to the sample. The systems are divided into two groups according to the used sample from \cite{2025MNRAS.538.1427P}: those with spectroscopic measurements (spectroscopic, SP) and those analyzed using photometry only (photometric, PH). The distributions of observation types (PH or SP), eclipse types (total or partial), and the ranges of orbital periods and primary masses for the sample used are summarized in Figure \ref{Fig:P-Mm-sample}. Pearson and Spearman correlation coefficients were computed independently for each group in order to determine whether the relation between $P$ and $M_{\rm m}$ behaves similarly in both subsets. The PH group exhibits a weaker correlation compared to the SP group, with Pearson and Spearman correlation coefficients of 0.609 and 0.656 for SP, and 0.445 and 0.632 for PH, respectively. The statistical properties of the two groups were assessed using Kolmogorov–Smirnov (KS; \citealt{smirnov1948table}) tests on both $M_{\rm m}$ and $P$. The distribution of $M_{\rm m}$ differs significantly between SP and PH (KS statistic = 0.317, $p < 0.001$). Similarly, the distribution of $P$ shows a significant difference (KS statistic = 0.316, $p < 0.001$). These results indicate a clear distinction between the two groups. An Analysis of Covariance (ANCOVA; \citealt{fisher1928statistical}) test was performed to evaluate whether the two groups share a common linear trend. The results show that the interaction term between Period and the data group (SP vs. PH) is not statistically significant ($F = 0.684$, $p = 0.409$), indicating that the slope of the mass-period relation does not differ significantly between the two groups. The main effect of Period is highly significant ($F = 142.822$, $p = 4.06 \times 10^{-29}$), while the effect of Group alone is not significant ($F = 1.323$, $p = 0.251$). Because of its higher intrinsic correlation and better-behaved residuals, the SP group was adopted as the reference subset for deriving the empirical mass-period relationship. Although the slopes of the mass-period relation do not differ significantly between the SP and PH groups according to the ANCOVA test, the SP subset exhibits a stronger correlation (Pearson = 0.609, Spearman = 0.656), a smaller scatter (RMSE = 0.269), and a higher coefficient of determination ($R^2 = 0.371$) compared to the PH group (RMSE = 0.405, $R^2 = 0.198$). These metrics demonstrate that the SP group provides a more reliable and better-constrained empirical calibration, justifying its use for deriving the final mass-period relation. Additionally, Kernel Density Estimates (KDE) were used to visualize the underlying structure of the dataset. One-dimensional KDEs of $M_{\rm m}$ and $P$ for both SP and PH subsets are shown in Figure \ref{Fig:P-Mm}d,e providing an assessment of the distribution shapes and the degree of overlap between the groups. A two-dimensional KDE in the orbital period-mass plane is presented in Figure \ref{Fig:P-Mm}f, highlighting the principal density ridge and identifying the region where the majority of systems are located. This probability map suppresses the influence of outliers and allows a robust visual identification of the dominant empirical trend. A linear regression was performed on the SP group using a least-squares model with a constant term. Confidence intervals for the slope and intercept were computed through the statistical framework implemented in statsmodels (\citealt{seabold2010statsmodels}), and the resulting best-fitting relation was accompanied by a prediction interval corresponding to a $90\%$ confidence level. The regression lines for both PH and SP groups, together with their respective shaded confidence bands, are displayed in Figure \ref{Fig:P-Mm}a,b. The agreement between the fitted line and the two-dimensional KDE structure in Figure \ref{Fig:P-Mm}f confirms that the model follows the intrinsic trend of the data rather than being driven by a small fraction of points.

The final empirical relationship derived for the more massive ($M_{\rm m}$) star in short period contact binaries is,

\begin{equation}
M_{\rm m} = (3.128 \pm 0.627)\,P + (0.142 \pm 0.240),
\label{eq:mm_p_relation}
\end{equation}
where $P$ and $M_{\rm m}$ are expressed in days and solar units, respectively. The parameters correspond to the coefficients obtained from the SP subset using the linear regression with the associated one–sigma uncertainties. Figure \ref{Fig:P-Mm}d, e, f, and a summarize the complete statistical procedure, including the group comparisons, one-dimensional KDEs, two-dimensional KDE map, ANCOVA behavior, and the final regression model.

The empirical relationship reported by \cite{2025MNRAS.538.1427P} is overplotted in panels~a and~b of Figure~\ref{Fig:P-Mm} for comparison. As evident from Figure~\ref{Fig:P-Mm}a, the final relation derived in this study from the SP data deviates from that reported by \cite{2025MNRAS.538.1427P}. In contrast, it closely matches the fit obtained using only the PH data in Figure~\ref{Fig:P-Mm}b. This suggests that when all data are included in the analysis, the PH sample exerts a dominant influence on the resulting relation. This behavior is likely driven, at least in part, by the larger representation of PH systems in the sample, which naturally increases their statistical weight in the combined fit. In addition, systematic differences in the determination of stellar masses from photometric modeling may contribute to the observed discrepancy between the SP-based and PH-based relations. Since mass estimates derived from light curve solutions depend on modeling assumptions and parameter degeneracies, any methodological differences can propagate into the inferred relation. Combining sub-samples without proper separation may affect the resulting empirical relation. Therefore, the SP and PH samples are treated separately in this study to assess their individual contributions to the derived $P$-$M_m$ relationship. Consequently, unlike \cite{2025MNRAS.538.1427P}, in which the empirical relation was fitted using a mixed PH and SP sample, with the PH sample dominating the fit, the revised empirical relation is derived exclusively from systems with SP masses.

The absolute parameters of the 10 target systems were computed using their orbital periods, mass ratios, mean fractional radii, and effective temperatures. The empirical $P$-$M_{\rm m}$ relation derived in this work was first used to estimate the mass of the more massive component. The mass of the less massive star was then computed from the mass ratio, adopting $M_2 = q \times M_1$ or $M_1 = M_2/q$ depending on whether $q < 1$ or $q > 1$.

The semi-major axis $a(R_{\odot})$ was derived from Kepler's third law in solar units, and the component radii from the light curve solution (Table \ref{Tab:lc-analysis}) were calculated as $R_{1,2} = r_{1,2} \times a$. Luminosities followed from $L_{1,2} = R_{1,2}^2 \times (T_i / T_\odot)^4$, and bolometric magnitudes were obtained using $M_{{\rm bol}_{1,2}} = M_{{\rm bol}\odot} - 2.5 \log L_{1,2}$, where a solar bolometric magnitude of $M_{\rm bol\odot} = 4.73$ (\citealt{2010AJ....140.1158T}) was adopted. The absolute magnitudes ($M_{V1,2}$) were estimated using bolometric corrections ($BC_{1,2}$) derived from \cite{flower1996transformations}.
Surface gravities were evaluated through $g_{1,2} = G_\odot\,M_{1,2}/R_{1,2}^2$ and given in logarithmic form. The orbital angular momentum was also computed using Equation \ref{eqJ0}.

\begin{equation}\label{eqJ0}
J_0 = \frac{q}{(1+q)^2} \sqrt[3]{\frac{G^2}{2\pi} M^5 P}.
\end{equation}

The final parameter estimates for the 10 target systems, along with the corresponding $1\sigma$ uncertainties, are presented in Table \ref{Tab:absolute}.

\begin{sidewaystable*}
\renewcommand\arraystretch{1.5}
\setlength{\tabcolsep}{3pt}
\caption{Absolute stellar parameters derived for the analyzed contact binaries.}
\centering
\footnotesize
\begin{tabular}{c c c c c c c c c c c}
\hline
Parameter & G1721 & G2161 & NO Leo & TIC 422347573 & V337 UMa & V369 Boo & V475 Ser & V640 Aur & V689 Vir & V702 Aur\\
\hline
$M_1(M_\odot)$ 	&	 0.401(139) 	&	 1.075(427) 	&	 0.331(116) 	&	 0.089(3) 	&	 0.467(135) 	&	 1.179(448) 	&	 0.314(123) 	&	 0.834(279) 	&	 0.588(206) 	&	 0.929(272) \\
$M_2(M_\odot)$ 	&	 1.063(424) 	&	 0.136(72) 	&	 1.052(422) 	&	 0.932(398) 	&	 1.341(480) 	&	 0.980(403) 	&	 0.892(390) 	&	 1.168(445) 	&	 1.018(415) 	&	 1.187(449) \\
$R_1(R_\odot)$ 	&	 0.636(80) 	&	 1.149(162) 	&	 0.603(93) 	&	 0.346(44) 	&	 0.793(84) 	&	 1.043(131) 	&	 0.522(85) 	&	 0.928(145) 	&	 0.721(104) 	&	 0.978(144) \\
$R_2(R_\odot)$ 	&	 0.994(121) 	&	 0.472(82) 	&	 1.011(135) 	&	 0.993(120) 	&	 1.288(138) 	&	 0.957(121) 	&	 0.835(124) 	&	 1.075(158) 	&	 0.924(128) 	&	 1.090(155) \\
$L_1(L_\odot)$ 	&	 0.426(128) 	&	 1.370(476) 	&	 0.191(72) 	&	 0.132(41) 	&	 0.558(147) 	&	 1.204(349) 	&	 0.143(58) 	&	 0.768(294) 	&	 0.351(122) 	&	 1.105(392) \\
$L_2(L_\odot)$ 	&	 0.877(258) 	&	 0.136(58) 	&	 0.361(117) 	&	 1.108(332) 	&	 1.160(304) 	&	 0.774(225) 	&	 0.322(119) 	&	 0.877(318) 	&	 0.440(149) 	&	 1.234(427) \\
$M_{bol1}(mag.)$ 	&	 5.658(286) 	&	 4.388(324) 	&	 6.529(346) 	&	 6.928(295) 	&	 5.363(253) 	&	 4.528(276) 	&	 6.841(370) 	&	 5.016(351) 	&	 5.865(323) 	&	 4.622(330) \\
$M_{bol2}(mag.)$ 	&	 4.872(280) 	&	 6.894(388) 	&	 5.836(305) 	&	 4.619(285) 	&	 4.569(253) 	&	 5.009(277) 	&	 5.962(342) 	&	 4.873(336) 	&	 5.621(317) 	&	 4.502(323) \\
$M_{V1}$(mag.) 	&	 5.726(280) 	&	 4.459(316) 	&	 6.879(326) 	&	 6.985(288) 	&	 5.476(243) 	&	 4.584(272) 	&	 7.190(346) 	&	 5.128(341) 	&	 6.081(310) 	&	 4.670(324)\\
$M_{V2}$(mag.) 	&	 4.985(270) 	&	 7.159(369) 	&	 6.473(279) 	&	 4.672(277) 	&	 4.770(240) 	&	 5.138(271) 	&	 6.394(315) 	&	 5.040(323) 	&	 5.980(296) 	&	 4.573(315)\\
$log\textit{(g)}_1$(cgs) 	&	 4.434(26) 	&	 4.348(30) 	&	 4.397(6) 	&	 4.309(92) 	&	 4.309(22) 	&	 4.473(37) 	&	 4.498(13) 	&	 4.424(1) 	&	 4.491(13) 	&	 4.425(7) \\
$log\textit{(g)}_2$(cgs) 	&	 4.470(46) 	&	 4.224(45) 	&	 4.451(37) 	&	 4.414(56) 	&	 4.346(45) 	&	 4.468(46) 	&	 4.545(37) 	&	 4.443(21) 	&	 4.514(36) 	&	 4.437(24) \\
$a(R_\odot)$ 	&	 2.114(242) 	&	 2.002(244) 	&	 2.059(238) 	&	 1.694(198) 	&	 2.706(277) 	&	 2.606(305) 	&	 1.729(217) 	&	 2.523(273) 	&	 2.109(243) 	&	 2.602(267) \\
$logJ_0$(cgs) 	&	 51.491(228) 	&	 51.057(280) 	&	 51.410(230) 	&	 50.809(119) 	&	 51.666(201) 	&	 51.885(241) 	&	 51.306(250) 	&	 51.820(221) 	&	 51.617(232) 	&	 51.869(209) \\
\hline																				
$A_V(mag.)$ 	&	 0.067(1) 	&	 0.114(1) 	&	 0.102(1) 	&	 0.069(2) 	&	 0.020(1) 	&	 0.031(1) 	&	 0.113(1) 	&	 0.297(1) 	&	 0.079(2) 	&	 0.110(2)\\
$BC_1$ 	&	 -0.068(6) 	&	 -0.071(8) 	&	 -0.350(20) 	&	 -0.057(7) 	&	 -0.113(10) 	&	 -0.056(4) 	&	 -0.349(24) 	&	 -0.112(10) 	&	 -0.216(13) 	&	 -0.048(6)\\
$BC_2$ 	&	 -0.113(10) 	&	 -0.265(19) 	&	 -0.637(26) 	&	 -0.053(8) 	&	 -0.201(13) 	&	 -0.129(6) 	&	 -0.432(27) 	&	 -0.167(13) 	&	 -0.359(21) 	&	 -0.071(8)\\
\hline
\end{tabular}
\label{Tab:absolute}
\end{sidewaystable*}

\vspace{0.6cm}
\section{Discussion and Conclusion}
This study presents the first comprehensive photometric and orbital-period investigation of ten short-period contact binaries based on a coordinated observing campaign carried out at six observatories. By combining homogeneous ground-based observations with available TESS photometry, the analysis provides a consistent set of photometric, geometric, and evolutionary parameters for all targets. Since these binaries are analyzed here for the first time in such a comprehensive framework, their detailed physical properties were not known a priori and therefore were not used as selection criteria. Instead, the targets were selected according to observational considerations, including their short orbital periods ($P<0.4$ d), the availability of sufficiently complete ground-based light curves, and the possibility of validating the photometric solutions using at least two independent photometric passbands whenever possible. Nevertheless, the resulting sample spans a broad range of physical characteristics, including a binary close to the short-period cutoff (V475 Ser), systems with extremely low mass ratios (G2161 and TIC 422347573), and binaries exhibiting different types of orbital period evolution.

\subsection{Implications of the Orbital Period Variations}
We investigated the orbital period variations of the 10 systems. Six systems show long-term increasing or decreasing trends, three of which also exhibit cyclic variations. In contrast, the remaining four systems show no significant period variation and appear essentially constant.

Mass transfer within the binary stars is the most likely mechanism responsible for the observed long-term increase or decrease in the orbital period. Under the assumption that both mass and angular momentum are conserved, the mass transfer rate can be evaluated from the following equation:
\begin{equation}
\frac{dM_1}{dt}= \frac{M_1M_2}{3P(M_1-M_2)}\,\frac{dP}{dt}.
\end{equation}

By combining the orbital period change rate $\mathrm{d}P/\mathrm{d}t$ of each target with the masses of the two components, we computed the mass-transfer rate $\dot{M}_{1}$ ($\mathrm{d}M_{1}/\mathrm{d}t$) for every system; the results are listed in Table \ref{tab:oc_coeff}. A positive value indicates that the primary star $M_{1}$ is gaining mass, whereas a negative value signifies that it is losing mass.

A long-term decreasing period is caused by mass transfer from the more massive star to the less massive star or angular momentum loss. We used $\tau_{\mathrm{th}} = \frac{G M_1^2}{R_1 L_1}$ to calculate the thermal timescale, obtaining $\tau_{\mathrm{th}}=2.99\times10^{7}\,\mathrm{yr}$ for NO Leo. The corresponding thermal-timescale mass-transfer rate for NO Leo is $M_1/\tau_{\mathrm{th}}=1.11\times10^{-8}\,M_\odot\,\mathrm{yr^{-1}}$. For NO Leo, the thermal mass-transfer rate is comparable to the value reported in Table \ref{tab:oc_coeff}, indicating that mass transfer is a plausible explanation for the observed long-term orbital-period decrease.

About G1721, G2161, V369 Boo, V475 Ser and V702 Aur, the long-term increasing trend of the period are due to mass transfer from the less massive star to the more massive one. Owing to the conservation of angular momentum, the orbital separation grows as mass is transferred, reducing the degree of contact. Consequently, the systems will evolve from their present contact configuration into a semi-detached or even detached state. This behaviour is known as the Thermal Relaxation Oscillation (TRO) model of contact binaries \citep{1979ApJ...231..502L, 1976ApJ...205..217F, 1977MNRAS.179..359R}. Long-term monitoring of these targets is therefore essential.

Two mechanisms are commonly invoked to explain cyclic changes in the orbital period: the light-travel time effect produced by a tertiary companion \citep{2010MNRAS.405.1930L,2010PASP..122..935E,2015NewA...41...17L,2016Ap&SS.361...63L} and magnetic activity within the binary \citep{2014ApJ...788...48S}. When evaluating the tertiary-companion hypothesis, we adopted the method described by \citet{2013AJ....145...39Z}. If the third body is assumed to move in a circular orbit, the following equations apply:
\begin{equation}
a_{12} \times\sin i' = {A_3}\times{c},
\end{equation}
\begin{equation}
f(m) = \frac{(M_3 \sin i')^{3}}{(M_1 + M_2 + M_3)^{2}} 
       = \frac{4\pi^{2}}{G {P_{3}}^{2}}\, (a_{12} \sin i')^{3},
\end{equation}
where $M_1$ and $M_2$ denote the masses of the two binary components, while $M_3$ is the mass of the third body.  
We derived the mass function $f(m)$ of the additional component for the three targets.  
The orbital separation between the tertiary and the central binary can be estimated from
$a_3 = \frac{(M_1+M_2)\,a_{12}}{M_3}$.
If the orbital inclination of the third body \(i'\) is the same as the binaries (\(i' = i\)), the mass and the distance of the tertiary companion are calculated. All parameters are listed in Table \ref{tabel:applegate}.

Magnetic activity provides an alternative explanation for the observed cyclic oscillations through changes in the gravitational quadrupole moment of one or both stars \citep{1992ApJ...385..621A}. The relative amplitude of the period modulation is $\Delta P/P=2\pi A_{3}/P_{3}$, where $A_{3}$ and $P_{3}$ denote the semi-amplitude and period of the cyclic oscillation, respectively.
By applying the following equation (\citealt{2002AN....323..424L}),
\begin{equation}
\frac{\Delta P}{P}=-9\,\frac{\Delta Q}{M a^{2}},
\label{eq:DQ}
\end{equation}
where $M$ denotes the mass of the corresponding binary component and $a$ is the semi-major axis of the binary orbit.
The change in the gravitational quadrupole moment $\Delta Q$, can then be computed for each star. 
The quadrupole momenta of the required variation of both components ($\Delta Q_{1}$ and $\Delta Q_{2}$) were determined and are tabulated in Table \ref{tabel:applegate}.  
The typical value is usually $10^{51}$\,--\,$10^{52}$\,g\,cm$^{2}$ for close binaries, and $\Delta Q = 10^{49}$\,g\,cm$^{2}$ for cataclysmic variables (\citealt{1999A&A...349..887L}).  The gravitational quadrupole moments calculated for G2161, NO Leo, and V702 Aur are not fully consistent with the values commonly reported for similar systems. Nevertheless, they are all of the order of $10^{50}$\,g\,cm$^{2}$, as found for DZ Psc (\citealt{2013AJ....146...35Y}), V1101 Her (\citealt{2017AJ....154..260P}), and V0474 Cam (\citealt{2018PASP..130f4201G}).

\begin{table*}
\centering
\caption{Parameters of the best-fitting periodic model applied to the O–C variations of G2161, NO Leo, and V702 Aur.}
\label{tabel:applegate}
\begin{tabular}{lcccccc}
\hline
Star & $a'\sin i'$  & $f(m)$ ($M_{\odot}$) & $M_3$ ($M_{\odot}$) & $a_3$ ($R_{\odot}$) & $\Delta Q_1$ (g\,cm$^2$) & $\Delta Q_2$ (g\,cm$^2$) \\
\hline
G2161    & 7.530(90)  & 18.76341(67239) & 28.050(1407) & 0.42(2) & $7.24\times10^{50}$ & $9.15\times10^{49}$ \\
NO Leo   & 0.217(31)  & 0.00020(9)  & 0.080(19) & 4.63(126) & $4.55\times10^{48}$ & $1.45\times10^{49}$ \\
V702 Aur & 3.622(454) & 0.64113(24105) & 2.685(825) & 3.60(120) & $2.81\times10^{50}$ & $3.59\times10^{50}$ \\
\hline
\end{tabular}
\end{table*}

\vspace{0.4cm}
\subsection{Evaluation of the Derived Physical Properties}
This work presents the first investigation of 10 contact binary stars based on both ground- and space-based photometric data. Light curve analysis indicates that the effective temperatures of the stellar components in the selected systems span from 4451 K to 5983 K. TIC 422347573 has the smallest temperature difference between its components (25 K), whereas G2161 exhibits the largest difference (723 K). Temperature differences ($\Delta T=|T_1 - T_2|$) for all systems are summarized in Table \ref{Tab:conclusion}, where the associated uncertainties were determined using standard propagation of the temperature measurement errors. Spectral category for each star was determined according to temperature-based criteria from \cite{2018MNRAS.479.5491E}, as listed in Table \ref{Tab:conclusion}.

The initial and final estimations of the component effective temperatures were carried out following standard procedures commonly used in MCMC and Monte Carlo simulations. The effective temperatures reported in Gaia DR3 and TIC correspond to the unresolved system and do not represent the temperature of an individual component. Given that contact binaries are expected to exhibit nearly equal component temperatures due to efficient energy transfer within the common envelope, both component temperatures are anticipated to lie close to this system value. We therefore adopt the catalog temperature as an initial estimate and assign it to the hotter component at the beginning of the modeling procedure. In the subsequent MCMC analysis, the temperatures were treated as free parameters to sample the posterior distribution and quantify parameter correlations. The exploration was restricted to $\pm50$ K for the hotter component and $\pm100$ K for the cooler component around the initial solution, thereby defining the prior ranges for the sampling. Within these bounds, the posterior distributions remained well confined and centered near the initial values, with the effective exploration corresponding to a few standard deviations around the solution. Furthermore, the final temperature of the hotter star was required to remain within approximately $\pm400$ K of the Gaia value, consistent with the empirical constraint studied by \cite{2025MNRAS.537.3160P}. Although the light curve morphology primarily constrains the temperature ratio, allowing limited variation of both temperatures ensures a realistic estimation of uncertainties without introducing significant degeneracy.

According to the photometric solution results, target stars are partial-eclipse binaries (Table \ref{Tab:lc-analysis}). According to previous studies, we know that total-eclipse contact binaries provide stronger geometric constraints on the light curve analysis, which allows the stellar parameters, and in particular the photometric mass ratio, to be determined with higher reliability (\citealt{2021AJ....162...13L,2023AA...672A.176P,2023ApJ...958...84K}). In contrast, the accuracy of photometric mass ratios in partial-eclipse systems has often been questioned because statistical comparisons between spectroscopic ($q_{\text{sp}}$) and photometric ($q_{\text{ph}}$) mass ratios generally show closer agreement in total-eclipse systems (\citealt{2021AJ....162...13L,2024RAA....24j5002S}). This raises the question of whether accurate photometric mass ratios can also be obtained in partial-eclipse binaries. The study by \cite{2024AJ....168..272P}, using a sample of contact binaries with both $q_{\text{sp}}$ and $q_{\text{ph}}$ measurements, shows that while total-eclipse systems follow a nearly one-to-one relationship between $q_{\text{sp}}$ and $q_{\text{ph}}$, some carefully modeled partial-eclipse systems exhibit a similar trend. Therefore, the \cite{2024AJ....168..272P} study concluded that with precise light curve modeling, for example, using MCMC or Monte Carlo simulations, partial-eclipse systems can also provide reasonably reliable photometric mass ratios. In this study, we practically employed three methods to estimate the mass ratio. First, to obtain an initial estimate of the mass ratio for each system, we used the $q$-search and Kouzuma methods. Then, the final mass ratio was determined using MCMC, which is consistent with the results of \cite{2024AJ....168..272P}.

The conventional classification of W UMa-type binaries distinguishes two subtypes: A and W. In A-type systems, the more massive component is hotter, whereas in W-type systems it is cooler (\citealt{1970VA.....12..217B}). According to \cite{2020MNRAS.492.4112Z}, A- and W-type contact binaries are associated with different evolutionary scenarios. These differences are manifested in their structural configuration, thermal behavior, angular momentum evolution, and mass-exchange processes (\citealt{qian2020contact}). Our photometric solutions, together with the absolute parameters estimated using the new empirical relationship derived in this work, show that three of the investigated systems belong to the A subtype, whereas the remaining binaries are classified as W-subtype systems (Table~\ref{Tab:conclusion}).

\begin{figure*}
\centering
\includegraphics[width=0.99\textwidth]{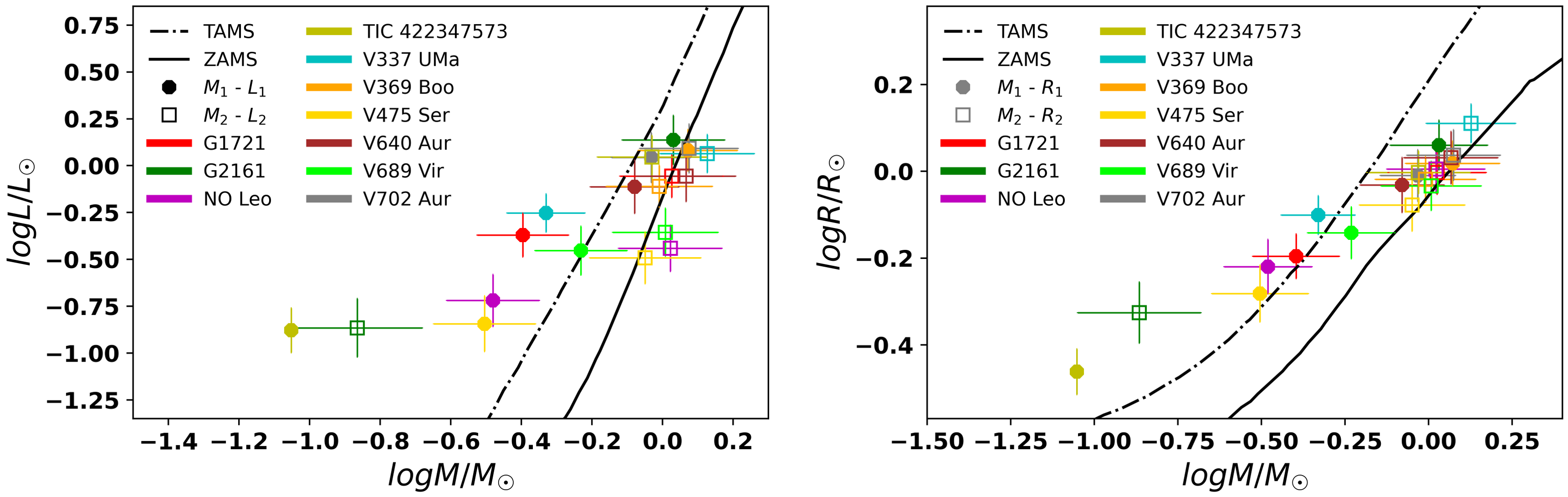}
\caption{Mass-Luminosity (left) and Mass-Radius (right) diagrams for the 10 systems analyzed in this study.}
\label{Fig:MLR}
\end{figure*}

Logarithmic Mass-Radius ($M$-$R$) and Mass–Luminosity ($M$-$L$) diagrams, constructed from the estimated absolute parameters, illustrate the evolutionary status of the target systems (Figure \ref{Fig:MLR}). Stellar components are plotted relative to the Zero-Age Main Sequence (ZAMS) and Terminal-Age Main Sequence (TAMS) boundaries defined by \cite{2000A&AS..141..371G}. In these diagrams, less massive companions generally appear closer to the TAMS, while the more massive stars lie nearer to the ZAMS. This does not necessarily imply that the less massive components evolve faster. In contact binaries, mass transfer and the resulting interaction between the components can alter their evolutionary paths, so that the evolutionary state of a component is not determined solely by its mass. The proximity of the less massive components to the TAMS therefore reflects the coupled evolutionary history of the binary rather than a shorter nuclear evolutionary timescale associated with their lower masses (\citealt{2006AcA....56..199S,2012AcA....62..153S}). It is important to recognize that the evolution of contact binaries is governed by a variety of complex processes, including the exchange of mass and angular momentum (\citealt{2005ApJ...629.1055Y}), which cause their evolutionary tracks to deviate from those of isolated single stars. Consequently, interpretations based on the single-star ZAMS and TAMS evolutionary tracks must be treated with caution.

Understanding how contact binaries progress toward coalescence requires close examination of systems with exceptionally small mass ratios. Among the targets analyzed in this work, G2161 and TIC 422347573 have mass ratios of $q = 0.127$ and $1/q = 0.095$, respectively, placing them among the low mass ratio contact binaries. These two low mass ratio systems exhibited a deep and sharp minimum in the $q$-search (Figure \ref{Fig:q}). TIC 422347573 yielded only a single minimum across the entire mass ratio search range, while G2161 system showed two minima. Modeling was performed for both observed minima; the solution utilizing the deeper minimum produced a notably superior synthetic light curve.

Assessing whether a contact binary remains stable or is likely to evolve toward orbital decay and eventual merger requires evaluating the balance between its spin and orbital angular momenta. A commonly used diagnostic for this purpose is the ratio $J_{spin}/J_{0}$, which provides insight into the dynamical configuration of the system. To compute this ratio for the two systems, we adopted the equations presented by \cite{2015AJ....150...69Y}:

\begin{equation}\label{eqJspin}
\frac{J_{spin}}{J_{0}}=\frac{1+q}{q}\big[(k_1 r_1)^2 + q (k_2 r_2)^2 \big],
\end{equation}
in this expression, $k_{1,2}$ represent the dimensionless gyration radii of the components, while $r_{1,2}$ denote their corresponding fractional radii.

The dynamical stability of G2161 and TIC 422347573, both identified as low mass ratio candidates, was evaluated using the ratio $J_{\rm spin}/J_0$, as defined in Equation \ref{eqJspin}. A system is considered dynamically unstable if this ratio exceeds $1/3$, while lower values indicate stability (\citealt{1980A&A....92..167H}). In these calculations, the dimensionless gyration radii were adopted from the \cite{Poroatal2026aa} study, with $k_1^2 = 0.0600(3)$ and $k_2^2 = 0.1800(8)$ for TIC 422347573, and $k_1^2 = 0.0601(2)$ and $k_2^2 = 0.1684(283)$ for G2161. TIC 422347573 and G2161 were found to have $J_{\rm spin}/J_0 = 0.096 \pm 0.001$ and $0.186 \pm 0.001$, respectively. Both ratios are below the critical limit, indicating that neither system exhibits signs of dynamical instability. These results further suggest that the spin angular momentum contributes a relatively small fraction of the total angular momentum, supporting the dynamical stability of both contact binaries.

\begin{table*}
\renewcommand\arraystretch{1.2}
\caption{Summary of the physical characteristics of the analyzed binaries, including the component temperature difference, the spectral category of each star, the contact-binary subtype, and the contact degree classification.}
\centering
\begin{center}
\footnotesize
\begin{tabular}{c c c c c}
\hline
System & $|\Delta T|$ (K) & Sp. category & Subtype & $f$ classification\\
\hline
G1721 & 240(56) & G3-G7 & W & Shallow \\
G2161 & 723(68) & G3-K1 & A & Medium \\
NO Leo & 460(52) & K2-K5 & W & Shallow \\
TIC 422347573 & 25(70) & G2-G1 & A & Shallow \\
V337 UMa & 326(59) & G7-K0 & W & Shallow \\
V369 Boo & 387(36) & G2-G8 & A & Shallow \\
V475 Ser & 156(67) & K2-K3 & W & Shallow \\
V640 Aur & 221(65) & G7-K0 & W & Shallow \\
V689 Vir & 341(55) & K0-K2 & W & Shallow \\
V702 Aur & 156(64) & G1-G3 & W & Shallow \\
\hline
\end{tabular}
\end{center}
\label{Tab:conclusion}
\end{table*}

\vspace{0.6cm}
\section*{Data Availability}
The ground-based observations and the complete set of extracted times of minima for all systems are available in the online supplementary material accompanying this paper.

\vspace{0.6cm}
\section*{Acknowledgments}
This manuscript, including the observation, analysis, and writing processes, was provided by the BSN project (\url{https://bsnp.info}). This work is the result of the Binary Systems Spring School (B3S), held in April 2025 at the Observatoire de Haute-Provence-OHP in the south of France, with the support of the French Astronomical Society (SAF) Double Stars Committee and funding provided by the Observatoire de Paris through the API Pro-Am (\url{https://gemini.obspm.fr}). Work by Kai Li was supported by the National Natural Science Foundation of China (NSFC) (No. 12273018) and by the Qilu Young Researcher Project of Shandong University. The authors wish to express their sincere gratitude to Kaloyan Penev, the developer of AutoWISP, for his support with the data-reduction stage of this work, and to Marc Ferrari, Director of OHP, for his support, as well as to Stéphane Favard for technical assistance during the observations at OHP. We acknowledge Shinjirou Kouzuma for making available the code implementing the derivative-based mass ratio estimation method. We are deeply grateful to Ehsan Paki and Elham Sarvari for their valuable assistance.

\bibliography{References}
\bibliographystyle{aasjournal}

\end{document}